\documentclass[twocolumn]{aastex631}

\newcommand{\dmunits}{\ensuremath{{\,{\rm pc}\,{\rm cm}^{-3}}}\xspace}
\newcommand{\rmunits}{\ensuremath{{{\rm rad}\,{\rm m}^{-2}}}\xspace}

\usepackage{color}
\usepackage{graphicx, epstopdf} 
\usepackage{gensymb}
\usepackage{natbib}
\usepackage{amsmath}
\usepackage{amsfonts}
\usepackage{amsthm}
\usepackage{xspace}
\usepackage{multirow}
\usepackage{url}

\received{June 16, 2023}
\revised{August 28, 2023}
\accepted{September 12, 2023}
\submitjournal{ApJ}

\shorttitle{Timing of AO327 MSPs}
\shortauthors{Lewis et al.}

\graphicspath{{./}{figures/}}

\begin{document}

\title{Discovery and Timing of Millisecond Pulsars with the Arecibo 327 MHz Drift-Scan Survey}

\correspondingauthor{Evan F.~Lewis}
\email{efl0003@mix.wvu.edu}

\author[0000-0002-2972-522X]{Evan F.~Lewis}
\affiliation{West Virginia University, Department of Physics and Astronomy, P. O. Box 6315, Morgantown, WV 26506, USA}
\affiliation{Center for Gravitational Waves and Cosmology, West Virginia University, Chestnut Ridge Research Building, Morgantown, WV 26506, USA}

\author{Timothy E.~E.~Olszanski}
\affiliation{West Virginia University, Department of Physics and Astronomy, P. O. Box 6315, Morgantown, WV 26506, USA}
\affiliation{Center for Gravitational Waves and Cosmology, West Virginia University, Chestnut Ridge Research Building, Morgantown, WV 26506, USA}

\author{Julia S. Deneva}
\affiliation{College of Science, George Mason University, 4400 University Dr, Fairfax, VA 22030, USA}

\author[0000-0003-1307-9435]{Paulo C.~C.~Freire}
\affiliation{Max-Planck-Institut fur Radioastronomie MPIfR, Auf dem Hugel 69, D-53121 Bonn, Germany}

\author[0000-0001-7697-7422]{Maura A. McLaughlin}
\affiliation{West Virginia University, Department of Physics and Astronomy, P. O. Box 6315, Morgantown, WV 26506, USA}
\affiliation{Center for Gravitational Waves and Cosmology, West Virginia University, Chestnut Ridge Research Building, Morgantown, WV 26506, USA}

\author[0000-0002-7261-594X]{Kevin Stovall}
\affiliation{National Radio Astronomy Observatory, 1003 Lopezville Rd., Socorro, NM 87801, USA}

\author[0000-0001-8640-8186]{Manjari Bagchi}
\affiliation{The Institute of Mathematical Sciences, CIT Campus, Taramani, Chennai 600113, India}
\affiliation{Homi Bhabha National Institute, Training School Complex, Anushakti Nagar, Mumbai 400094, India}

\author[0000-0003-0669-865X]{Jose G.~Martinez}
\affiliation{Max-Planck-Institut fur Radioastronomie MPIfR, Auf dem Hugel 69, D-53121 Bonn, Germany}

\author[0000-0002-8509-5947]{Benetge B.~P.~Perera}
\affiliation{Arecibo Observatory, University of Central Florida, HC3 Box 53995, Arecibo, PR 00612, USA}

\begin{abstract}

We present the discovery and timing solutions of four millisecond pulsars (MSPs) discovered in the Arecibo 327 MHz Drift-Scan Pulsar Survey. Three of these pulsars are in binary systems, consisting of a redback (PSR J2055+1545), a black widow (PSR J1630+3550), and a neutron star--white dwarf binary (PSR J2116+1345). The fourth MSP, PSR J2212+2450, is isolated. We present the multiyear timing solutions as well as polarization properties across a range of radio frequencies for each pulsar. We perform a multiwavelength search for emission from these systems and find an optical counterpart for PSR~J2055+1545 in Gaia DR3, as well as a gamma-ray counterpart for PSR~J2116+1345 with the Fermi-LAT telescope. Despite the close colocation of PSR~J2055+1545 with a Fermi source, we are unable to detect gamma-ray pulsations, likely due to the large orbital variability of the system. This work presents the first two binaries found by this survey with orbital periods shorter than a day; we expect to find more in the 40\% of the survey data that have yet to be searched.

\end{abstract}

\section{Introduction} 
\label{sec:intro}
Recycled pulsars are rapidly rotating neutron stars (NSs) with rotation periods shorter than 100 ms, believed to have been \textit{spun up} through their interactions with a low-mass star in a binary system \citep{acr+82}. The accretion of stellar material onto the pulsar leads to their high spin frequencies through angular momentum transfer, and neutron stars undergoing active accretion have been detected through their X-ray emission as low-mass X-ray binaries \citep{tv06}. Millisecond pulsars (MSPs) in particular are the shortest-period pulsars, typically defined as pulsars with rotation periods of $P < 30$ ms. Most recycled pulsars and MSPs retain binary companions; of the 695 pulsars in the ATNF pulsar catalog\footnote{Version 1.70; \url{https://www.atnf.csiro.au/research/pulsar/psrcat/}} with $P < 100$ ms, 347 pulsars (50\%) have confirmed binary companions. In addition to serving as testbeds of binary evolution, such binary systems can be used for a variety of fundamental physics experiments, including those that test general relativity and alternative theories of gravity, by measuring relativistic effects in the pulsar's orbit and in the propagation of the radio pulses to Earth (see \citealt{ksv+21} and references therein, and more generally \citealt{bbc+15}).

Of particular interest for considerations on stellar evolution are \textit{spider} pulsars, which are characterized by the ablation of material from their stellar companions \citep{mser13}. These binary MSPs are typically broken down into two subcategories: black widows with low-mass ($\lesssim 0.05 \, {M_\odot}$) companions \citep{fst88} and redbacks with higher-mass ($0.2-0.4 \, {M_\odot}$), often nondegenerate, stellar companions \citep{asr+09}. Pulsed radio emission from these pulsars is often difficult to detect due to eclipsing, scattering, and other effects of ionized material in their companion winds \citep{Fermi2PC}, apart from their high accelerations. Gamma-ray emission is not affected by the intervening material, allowing for the discovery of many spider systems through their association with gamma-ray sources, most of which have been discovered by the Fermi Large Area Telescope (LAT; \citealt{FermiLAT}). Radio pulsations from 24 black widow and 10 redback pulsars\footnote{\url{http://astro.phys.wvu.edu/GalacticMSPs/}} have been discovered through the follow-up of Fermi sources (\citealt{ssc+19, cms+21, ssc+22, TRAPUM}). 

Even beyond the spider population, many MSPs as well as young pulsars are known to be gamma-ray emitters, and gamma-ray pulsars tend to have relatively stable fluxes over time and curved spectra \citep{Fermi2PC}. Pulsar-like Fermi sources provide excellent candidates for radio searches \citep{rap+12}, and 
these targeted searches continue to make new discoveries (e.g. \citealt{drc+21,TRAPUM}). 

In parallel to targeted searches, wide-area untargeted surveys continue to be useful. The Arecibo Observatory 327 MHz Drift-Scan Pulsar Survey (AO327) is a survey of the entire sky visible to the former Arecibo 305 m telescope (declinations ranging from $-1^\circ$ to $38^\circ$) at 327 MHz \citep{dsm+13}. The survey began in 2010 and positioned the telescope at a fixed azimuthal and zenith angle, allowing Earth to rotate the beam across the sky. Due to the low observing frequency, AO327 is relatively more sensitive to pulsars with low dispersion measures (DMs; or integrated free electron densities along the line of sight), since high-DM signals tend to be much more scattered, lowering our sensitivity. Pulsars generally tend to have steep spectral indices, meaning they have higher flux densities at lower frequencies; we thus expect AO327 to also be more sensitive to steeper-spectrum pulsars. Without tracking, a source has a maximum transit time through the beam of approximately one minute, meaning that AO327 preserves sensitivity to tight, relativistic binaries with short orbital periods.

Since its inception, AO327 has discovered 96 pulsars\footnote{\url{http://www.naic.edu/~deneva/drift-search/}}, including 10 MSPs \citep{dsm+13,mgf+19}. Approximately 60\% of the total data volume of AO327 has been searched, and we expect many more pulsar discoveries in the remaining data that are currently being processed (T.~Olszanski et al. 2023, in preparation).  Notable AO327 discoveries include the first strongly asymmetric double neutron star system (DNS) PSR~J0453+1559, for which the companion is the lightest known in such systems \citep{msf+15}; if it is an NS, it would be the lowest-mass NS with an accurately determined mass \citep{tj19}. A second DNS found in this survey, PSR~J1411+2551, is one of the lightest known 
\citep{msf+17}. Another notable discovery was PSR~J2234+0611 \citep{dsm+13}: its eccentric orbit ($e = 0.11$) and helium white dwarf (He WD) companion \citep{aks+16} make it a member of a rare class of eccentric MSP - He WD systems. Its high orbital eccentricity and timing precision allowed a precise determination of the masses and full 3D position, orbital orientation, and systemic motion \citep{sfa+19}.

Of the 16 recycled pulsars discovered by AO327, which include 10 MSPs, three have such precise timing that they are now being observed as part of the North American Nanohertz Observatory for Gravitational Waves pulsar timing array project in an effort to detect nanohertz gravitational waves \citep{NG15}. These discoveries highlight one of the main advantages of wide, shallow surveys like AO327-- the discovery of bright objects. In addition to their use in pulsar timing arrays, the long-term timing of bright binary pulsars can be used for the precise detection of small relativistic effects in the orbit, which allows for the measurement of NS masses and tests of theories of gravity. 

In this paper, we present the discovery and follow-up timing of four MSPs found by the AO327 survey, and their continued follow-up observations with the Giant Meterwave Radio Telescope (GMRT; \citealt{GMRT}). Section~\ref{sec:obs} details the observations with both observatories as well as the data reduction and timing analysis, and we describe the timing solutions, including binary parameters, for each system in Section~\ref{sec:timing}. In Section~\ref{sec:counterparts} we describe the search for counterparts in optical, X-ray, and gamma-ray wavelengths. We discuss the results in Section~\ref{sec:disc}, and conclude in Section~\ref{sec:conclusion}.

\section{Observations and Data Analysis}
\label{sec:obs}

\begin{figure*}
    \centering
    \includegraphics[width=\textwidth]{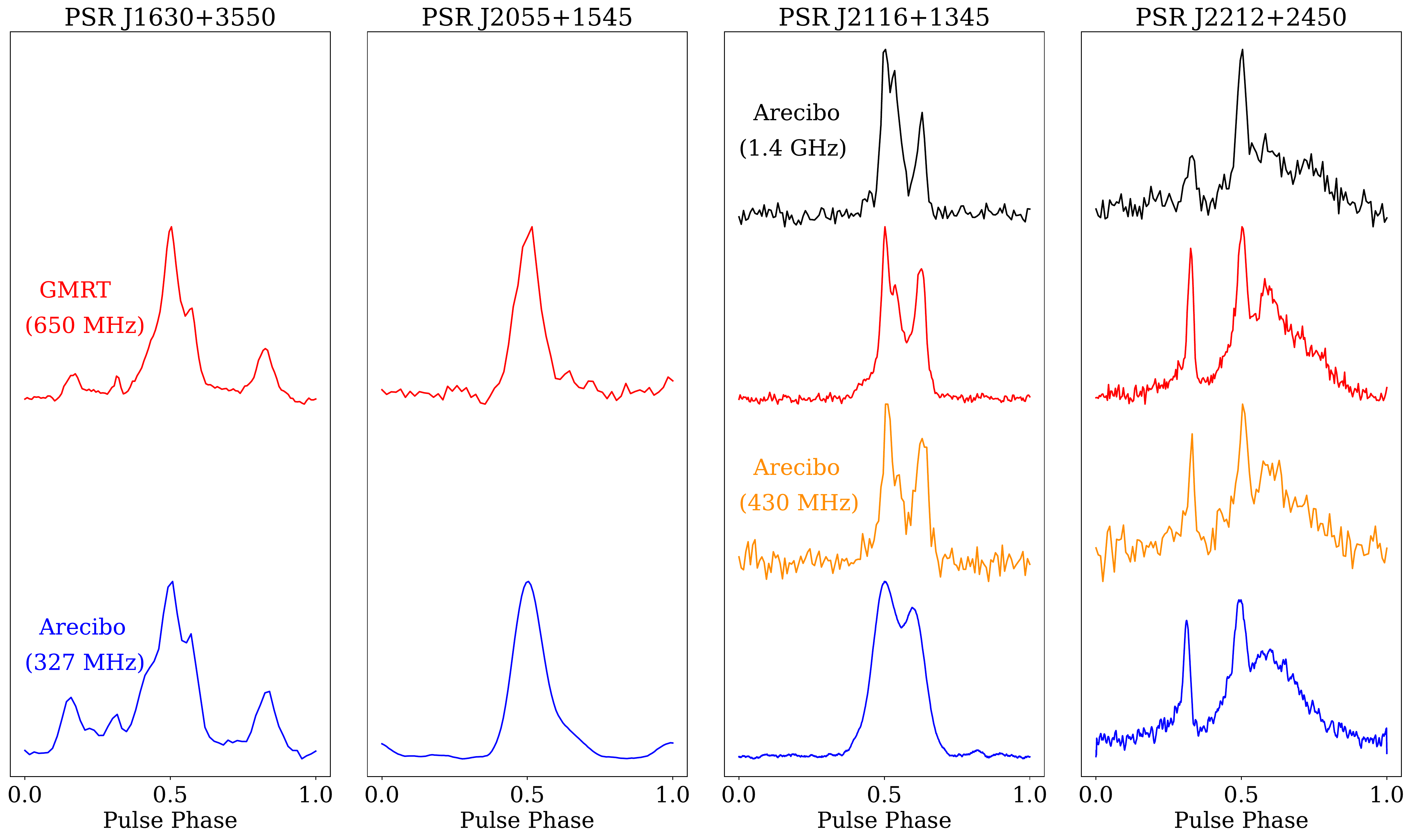}
    \caption{Composite pulse profiles of the four MSPs, with the colors indicating different observatories and observing frequencies. Profiles are centered at a pulse phase of 0.5 for clarity. From bottom to top-- blue: Arecibo, 327 MHz; orange: Arecibo, 430 MHz; red: GMRT, 650 MHz; and black: Arecibo, 1.4 GHz. Each profile has been normalized, and is vertically offset from one another for clarity. The disparities in the pulsar flux and number of observations at each frequency cause some profiles to have a lower S/N than others.}
    \label{fig:profs}
\end{figure*}

The discovery and initial follow-up observations of each pulsar were taken in search mode using the Arecibo 305 m dish, the 327 MHz receiver, and the Puerto Rico Ultimate Pulsar Processing Instrument (PUPPI) backend. These observations have a bandwidth of 68.7 MHz split into 2816 frequency channels, with a sampling time of 82 $\mu$s, as described in \citet{dsm+13}. A basic binary model for each such system was found by fitting the barycentric spin periods measured through the initial timing observations, after which we performed timing observations and data reduction following the same process as described in \citet{mgf+19}. We created composite template profiles for each pulsar by averaging together high signal-to-noise ratio (S/N) detections, and used the \textsc{presto}\footnote{\url{https://github.com/scottransom/presto}} \citep{smr01} software to fit Gaussian components to these summed profiles. Using these standard profiles, we generated the times of arrival (TOAs) by cross-correlating the observed pulse profiles with our template. We created separate template profiles for each frequency at which we observed the pulsars. At each epoch, we split the Arecibo data into two frequency sub-bands to create the TOAs.

In addition to the 327 MHz observations, we also obtained brief Arecibo observations of PSRs J2116+1345 and J2212+2450 at 430 MHz and 1.4 GHz. These data were coherently de-dispersed and folded using a basic timing model for each pulsar. The 1.4 GHz observations were taken with the $L$-band wide receiver, with a bandwidth of 800 MHz split into 512 frequency channels. The 430 MHz observations were taken with the 430 MHz receiver, with a bandwidth of 87.5 MHz split into 56 frequency channels.

We began each Arecibo observation with an observation of the noise diode, and observed the pulsars using full Stokes parameters. We calibrated the profiles using the \textsc{psrchive} software suite,\footnote{\url{https://psrchive.sourceforge.net}} and averaged together the best detections of each pulsar to create high S/N composite profiles. We also used the \texttt{rmfit} tool in \textsc{psrchive} to find the Faraday rotation measure (RM) which maximizes the detection of linear polarization in the composite profiles. We searched RMs from $-$200 to 200 \rmunits, using iterative position angle refinement to find the RM at which the flux of the linear polarization peaks. We present the errors reported by \texttt{rmfit} in Section~\ref{sec:timing}, noting that the RM measurements are likely more inaccurate than suggested by these error bars. 

We obtained timing observations of these four pulsars using the upgraded GMRT (uGMRT; \citealt{uGMRT}) wideband receiver and backend, with a central frequency of 650 MHz and a 200 MHz bandwidth. Using a phased array, we took coherently de-dispersed data with 10.24 $\mu$s time resolution and 512 frequency channels for precise TOA generation; we simultaneously took data at the same frequency without coherent de-dispersion using 81.92 $\mu$s time resolution and 4096 frequency channels in order to measure scintillation properties. All of these observations were taken with Stokes-$I$ total intensity, so no polarization information was recorded. 

Due to data availability issues, we were unable to incorporate the full year of GMRT data from each pulsar. This paper includes the subset of GMRT data that is currently available to us. For this initial study, we generate the TOAs by averaging over all of the frequency channels, instead of splitting the data up into multiple bands and creating sub-banded TOAs. Future studies may allow for a more thorough spectral analysis of these pulsars, including testing the ability of sub-banded TOAs to increase the timing precision of these pulsars.

\begin{figure*}
    \centering
\includegraphics[width=\textwidth]{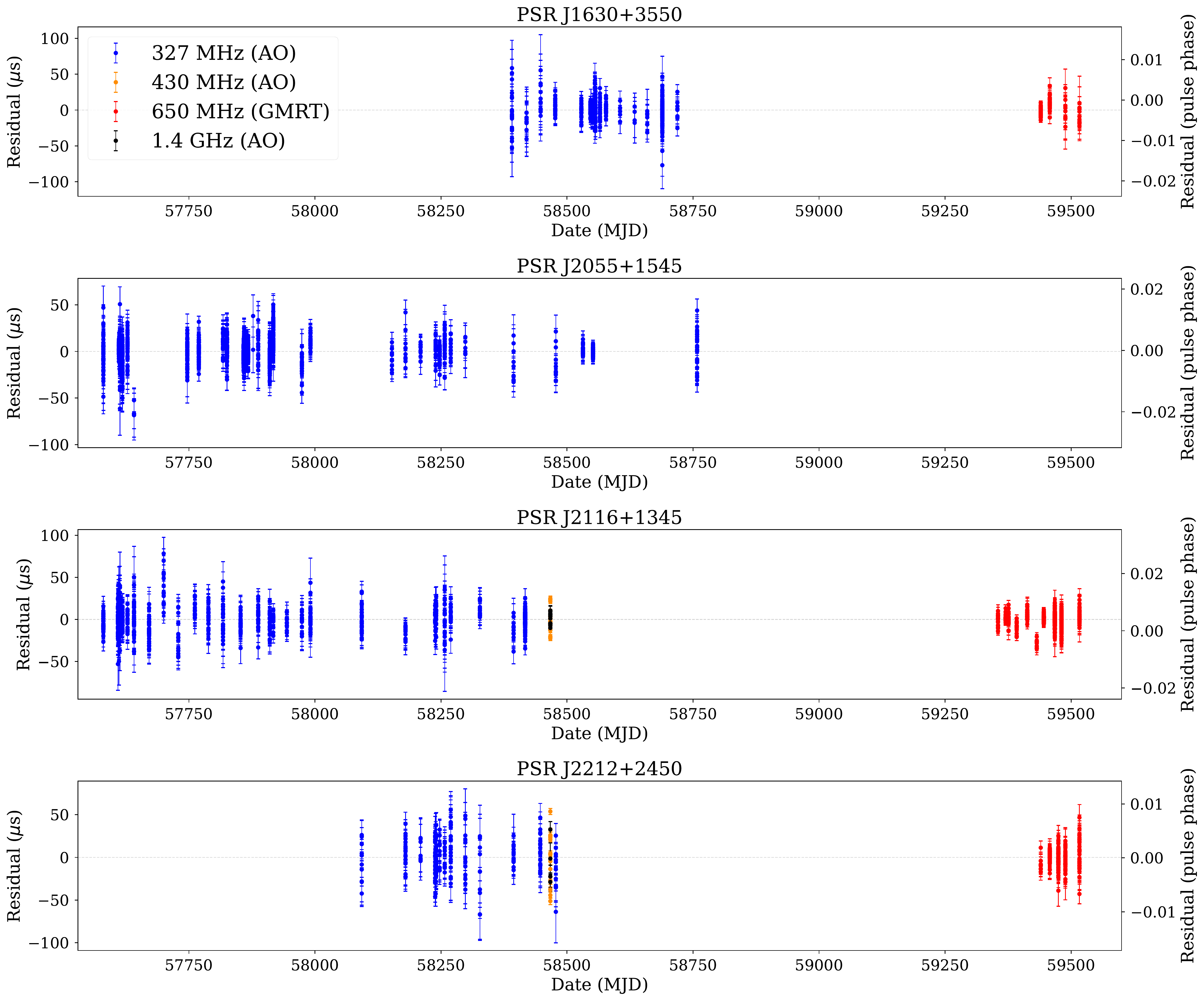}
    \caption{The TOA residuals as a function of observing date for the four pulsars in this paper, with 1$\sigma$ error bars. Blue: Arecibo, 327 MHz; orange: Arecibo, 430 MHz; red: GMRT, 650 MHz; and black: Arecibo, 1.4 GHz.}
    \label{fig:res_date}
\end{figure*}
\begin{figure*}
    \centering
\includegraphics[width=\textwidth]{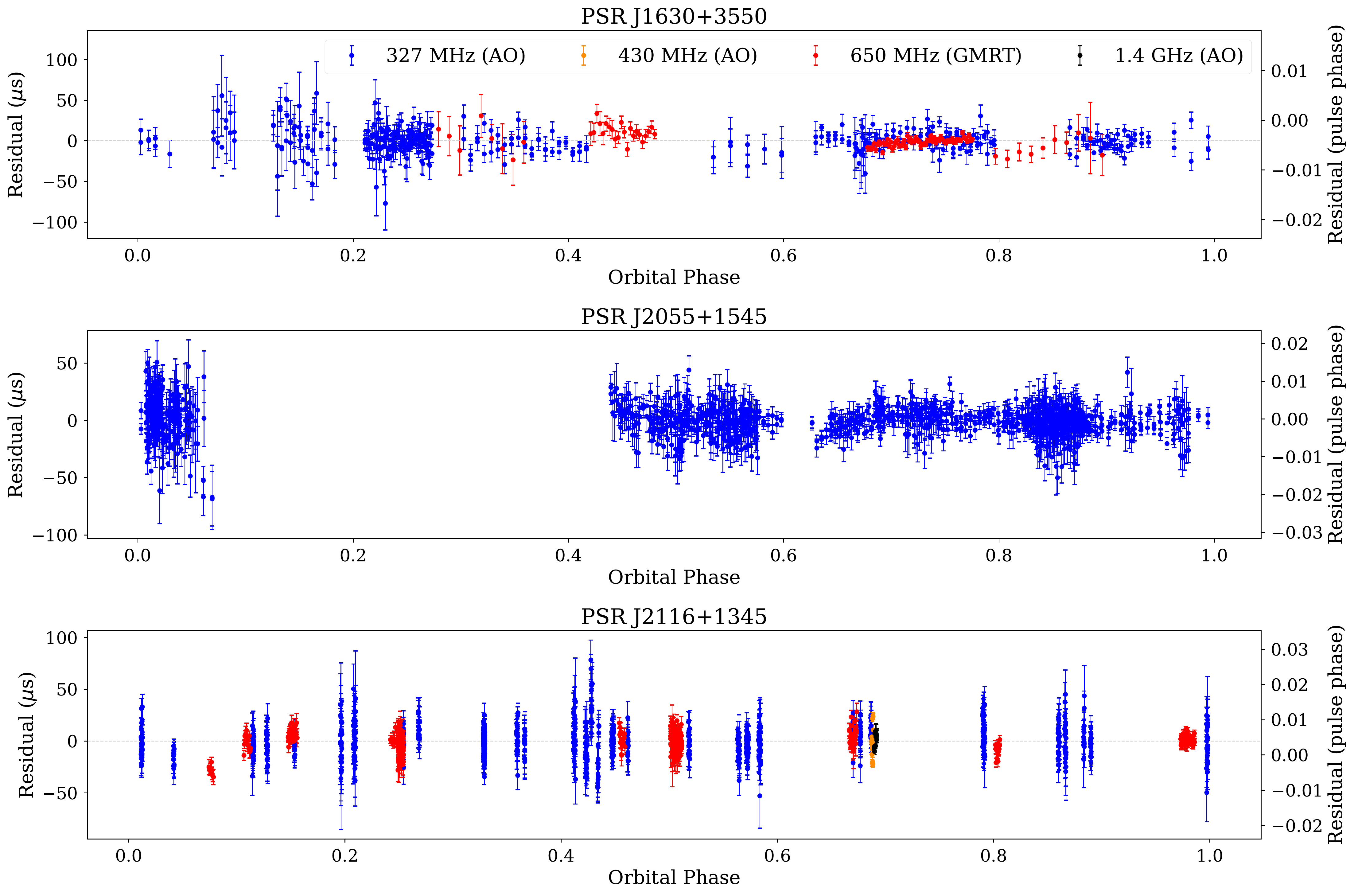}
    \caption{The TOA residuals as a function of orbital phase for the three binary pulsars in this paper, with 1$\sigma$ error bars. Blue: Arecibo, 327 MHz; orange: Arecibo, 430 MHz; red: GMRT, 650 MHz; and black: Arecibo, 1.4 GHz. An orbital phase of 0.25 corresponds to a superior conjunction.}
    \label{fig:res_orb}
\end{figure*}

\begin{deluxetable*}{cccc}
\tablewidth{\textwidth} 
\label{tab:ELL_solns}
\tablecaption{Timing parameters for pulsars using the ELL1 binary model}
\tablehead{& \colhead{PSR} & \colhead{J1630+3550} & \colhead{J2116+1345}} 
\startdata
\multirow{6}*{Data} & Data span (yr) & 3.1 & 14.4 \\
& Start epoch (MJD) & 58390 & 54682 \\
& End epoch (MJD) & 59516 & 59945 \\
& Timing epoch (MJD) & 58953 & 57628 \\
& rms post-fit residuals ($\mu$s) & 7.9 & 9.3\tablenotemark{a}\\
& Number of TOAs & 467 & 1258\tablenotemark{a} \\
\hline
\multirow{8}*{Pulsar parameters} & 
Right Ascension, $\alpha$ (J2000) & $16^{\rm h}\, 30^{\rm m}\, 35\, \fs94895(6)$ & $21^{\rm h}\, 16^{\rm m}\, 49\, \fs76983(7,-41)$ \\
& Declination, $\delta$ (J2000) & $+35\arcdeg\, 50\arcmin\, 42\, \farcs477(2)$ & $+13\arcdeg\, 45\arcdeg\, 21\, \farcs185(7,-5)$ \\
& Right ascension proper motion, $\mu_{\alpha}$ (mas yr$^{-1}$) & $-$22(7) & 7.9(2.3,$-$0.2) \\
& Declination proper motion, $\mu_{\delta}$ (mas yr$^{-1}$) & 33(29) & 7.7(0.6,$-$3.8) \\
& Spin frequency, $\nu$ (Hz) & 309.6784147487(2) &  450.752582185974(50,$-$9) \\
& Spin frequency derivative, $\dot{\nu}$ (Hz s$^{-1}$) & $-$2.163(5) $\times 10^{-15}$ & $-$5.369(1,$-$14) $\times 10^{-16}$\\
& Dispersion measure (pc cm$^{-3}$) & 17.4551(6) &  30.2899(1)\tablenotemark{a} \\ 
& Dispersion measure derivative (pc cm$^{-3}$ yr$^{-1}$) & 0.0028(5) & $-$9.6(8) $\times 10^{-4}$\tablenotemark{a} \\
\hline
\multirow{7}*{Binary parameters} & Binary model & ELL1 & ELL1 \\
& Orbital period, $P_b$ (days) & 0.315863166(1) & 4.81532204043(1,$-$1670) \\
& Projected semimajor axis of orbit, A1 (lt-s) & 0.036125(1) & 3.925461(8,$-$4) \\
& Epoch of ascending node, TASC (MJD) & 58355.096250(2) & 57582.3563765(33,$-$6) \\
& $\epsilon_1 \equiv e\rm{sin\omega}$ & $-$3(6) $\times 10^{-5}$ & $-$9.8(5.0,$-$1.6) $\times 10^{-6}$ \\
& $\epsilon_2 \equiv e\rm{cos\omega}$ & $-$4.2(7) $\times 10^{-4}$ & $-$9.0(3.2,$-$2.1) $\times 10^{-6}$ \\
\hline
\multirow{14}*{Derived parameters} & Spin period, $P$ (s) &  0.003229156287213(2) &  0.0022185119720233(3) \\
& Observed spin period derivative, $\dot{P}$ ($10^{-21}$ s s$^{-1}$) & 22.55(5) & 2.643(7) \\
& Spin-down luminosity, $\dot{E}$ $(10^{34}$ erg s$^{-1}$) & 2.64 & 0.95 \\
& Inferred surface dipole magnetic field, $B_{\rm{surf}}$ $(10^{8}$ G) & 2.7 & 0.77 \\
& Characteristic age, $\tau_c$ (Gyr) & 2.3 & 13.3 \\
& Minimum/median companion mass (${M_\odot}$) & 0.0098/0.0113 & 0.19/0.22 \\
& Orbital eccentricity, $e$ & 0.00042(7) & 0.000013(4) \\
& Longitude of Periastron, $\omega \; (^{\circ})$ & 184(9) & 230(20) \\
& Galactic latitude, $l \; (^{\circ})$ & 57.88 & 64.15 \\
& Galactic longitude, $b \; (^{\circ})$ & 43.08 & $-$23.82 \\
& DM distance (kpc); NE2001 & 1.1 & 1.8\tablenotemark{a} \\
& DM distance (kpc); YMW16 & 1.6 & 2.7\tablenotemark{a} \\
& Transverse velocity, $\rm{v}_{\rm{t}}$ (km s$^{-1}$); NE2001 & 210(140) & 100(40) \\
& Transverse velocity, $\rm{v}_{\rm{t}}$ (km s$^{-1}$); YMW16 & 300(200) & 140(60) \\
\hline
\enddata
\tablecomments{Timing parameters for the black widow pulsar PSR J1630+3550 and the NS-WD binary PSR J2116+1345, using the ELL1 binary model. The 1$\sigma$ error bars are denoted with parenthesis on the last significant digit. For J1630, the parameters are all taken from the radio timing solution. For J2116, values come from the Fermi solution unless otherwise noted, and asymmetric error bars come from the maximum likelihood estimation described in Section~\ref{sec:fermi}. The error bars on the derived parameters are calculated using the larger of the error bars on the measured parameters. Table~\ref{tab:Pdots} compares the observed and intrinsic spin-down rates for each pulsar. Errors in transverse velocity are quoted assuming 30\% uncertainties in the DM-based distance estimate.}
\tablenotetext{a}{Taken from the radio timing solution.}
\end{deluxetable*}
\begin{deluxetable*}{cccc}
\tablewidth{\textwidth} 
\label{tab:noELL_solns}
\tablecaption{Timing parameters for pulsars not using the ELL1 binary model}
\tablehead{ &
\colhead{PSR} & \colhead{J2055+1545} & \colhead{J2212+2450}
} 
\startdata
\multirow{6}*{Data} & Data span (yr) & 3.23 & 3.9 \\
& Start epoch (MJD) & 57580 & 58092 \\
& End epoch (MJD) & 58758 & 59516 \\
& Timing epoch (MJD) & 58169 & 58804 \\
& rms post-fit residuals ($\mu$s) & 9.53 & 19.6 \\
& Number of TOAs & 1209 & 384 \\
\hline
\multirow{9}*{Pulsar parameters} & Right Ascension, $\alpha$ (J2000) & $20^{\rm h}\, 55^{\rm m}\, 47\, \fs83166(4)$ & $22^{\rm h}\, 12^{\rm m}\, 27\, \fs574(1)$ \\
& Declination, $\delta$ (J2000) & $+15\arcdeg\, 45\arcmin\, 21\, \farcs2164(6)$ & $+24\arcdeg\, 50\arcmin\, 36\, \farcs80(1)$ \\
& Right ascension proper motion, $\mu_{\alpha}$ (mas yr$^{-1}$) & 3.3(7) & 8.5(6.5) \\
& Declination proper motion, $\mu_{\delta}$ (mas yr$^{-1}$) & $-$2.2(6) & 23(11) \\
& Spin frequency, $\nu$ (Hz)& 463.17144393614(5) & 255.917115508(2)\\
& Spin frequency derivative, $\dot{\nu}$ (Hz s$^{-1}$) & $-$2.096(3) $\times 10^{-15}$ & $-$1.3(4) $ \times 10^{-16}$\\
& Spin frequency second derivative, $\ddot{\nu}$ (Hz s$^{-2}$) & $-$1.2(1) $\times 10^{-24}$ & \nodata \\
& Dispersion measure (pc cm$^{-3}$) & 40.42296(6) & 25.2113(4) \\ 
& Dispersion measure derivative (pc cm$^{-3}$ yr$^{-1}$) & $-$3.5(5) $\times 10^{-4}$ & \nodata \\
\hline
\multirow{12}*{Binary parameters} & Binary model & BTX & \nodata \\
& Projected semi-major axis of orbit, A1 (lt-s) & 0.5996750(9) & \nodata \\
& Epoch of periastron, T$_0$ (MJD) & 57580.2997410(2) & \nodata \\
& FB0 (Hz) & 5.76612182(4) $\times 10^{-5}$ & \nodata \\
& FB1 (Hz s$^{-1}$) & 9.2(2) $\times 10^{-18}$ & \nodata \\
& FB2 (Hz s$^{-2}$) & $-$2.44(7) $\times 10^{-24}$ & \nodata \\
& FB3 (Hz s$^{-3}$) & 5.2(2) $\times 10^{-31}$ & \nodata \\
& FB4 (Hz s$^{-4}$) & $-$9.0(3) $\times 10^{-38}$ & \nodata \\
& FB5 (Hz s$^{-5}$) & 1.18(3) $\times 10^{-44}$ & \nodata \\
& FB6 (Hz s$^{-6}$) & $-$1.05(3) $\times 10^{-51}$ & \nodata \\
& FB7 (Hz s$^{-7}$) & 5.7(1) $\times 10^{-59}$ & \nodata \\
& FB8 (Hz s$^{-8}$) & $-$1.42(3) $\times 10^{-66}$ & \nodata \\
\hline
\multirow{12}*{Derived parameters} & Spin period, $P$ (s) & 0.0021590277489946(2) & 0.00390751512659(3) \\
& Observed spin period derivative, $\dot{P}$ ($10^{-21}$ s s$^{-1}$) & 9.772(8) & 2.0(6) \\
& Spin-down luminosity, $\dot{E}$ ($10^{34}$ erg s$^{-1}$) & 3.83 & 0.13 \\
& Surface magnetic field, B ($10^{8}$ G) & 1.47 & 0.89 \\
& Characteristic age, $\tau_c$ (Gyr) & 3.5 & 31 \\
& Minimum/median companion mass (${M_\odot}$) & 0.24/0.29 & \nodata \\
& Galactic latitude, $l \; (^{\circ})$ & 62.54 & 82.96 \\
& Galactic longitude, $b \; (^{\circ})$ & $-$18.57 & $-$25.43 \\
& DM distance (kpc); NE2001 & 2.4 & 1.59 \\
& DM distance (kpc); YMW16 & 3.7 & 2.09 \\
& Transverse velocity, $\rm{v}_{\rm{t}}$ (km s$^{-1}$); NE2001 & 46(16) & 180(100) \\
& Transverse velocity, $\rm{v}_{\rm{t}}$ (km s$^{-1}$); YMW16 & 69(24) & 240(130) \\
\hline
\enddata
\tablecomments{Timing parameters for the redback pulsar PSR J2055+1545 and the isolated MSP PSR J2212+2450. The 1$\sigma$ error bars are denoted with parenthesis on the last significant digit. J2055 uses the BTX model with eccentricity and $\omega$ set to 0. We notate the orbital frequency as FB0, and higher numbers represent its respective derivative. The proper motion values and errors for J2055 are taken from its Gaia counterpart. J2212 has no binary companion. Table~\ref{tab:Pdots} compares the observed and intrinsic spin-down rates for each pulsar. Errors in transverse velocity are quoted assuming 30\% uncertainties in the DM-based distance estimate.}
\end{deluxetable*}

\section{Radio Timing}
\label{sec:timing}
In this paper, we present the timing solutions derived using the entirety of the Arecibo data set as well as the GMRT data available to us. Each pulsar has a solution that spans at least 3 yr, with the timing solution for PSR J2116+1345 spanning over 14 yr, with the addition of the Fermi data set. The composite radio pulse profiles used to generate the TOAs at each frequency are shown in Figure~\ref{fig:profs}, and Tables~\ref{tab:ELL_solns} and~\ref{tab:noELL_solns} list the entirety of each pulsar's timing solution. 
We derived these timing solutions using the DE440 solar system ephemeris, except for the Fermi solution for J2116+1345, which uses the DE421 ephemeris.
  
\begin{figure}
    \centering
    \includegraphics[width=\linewidth]{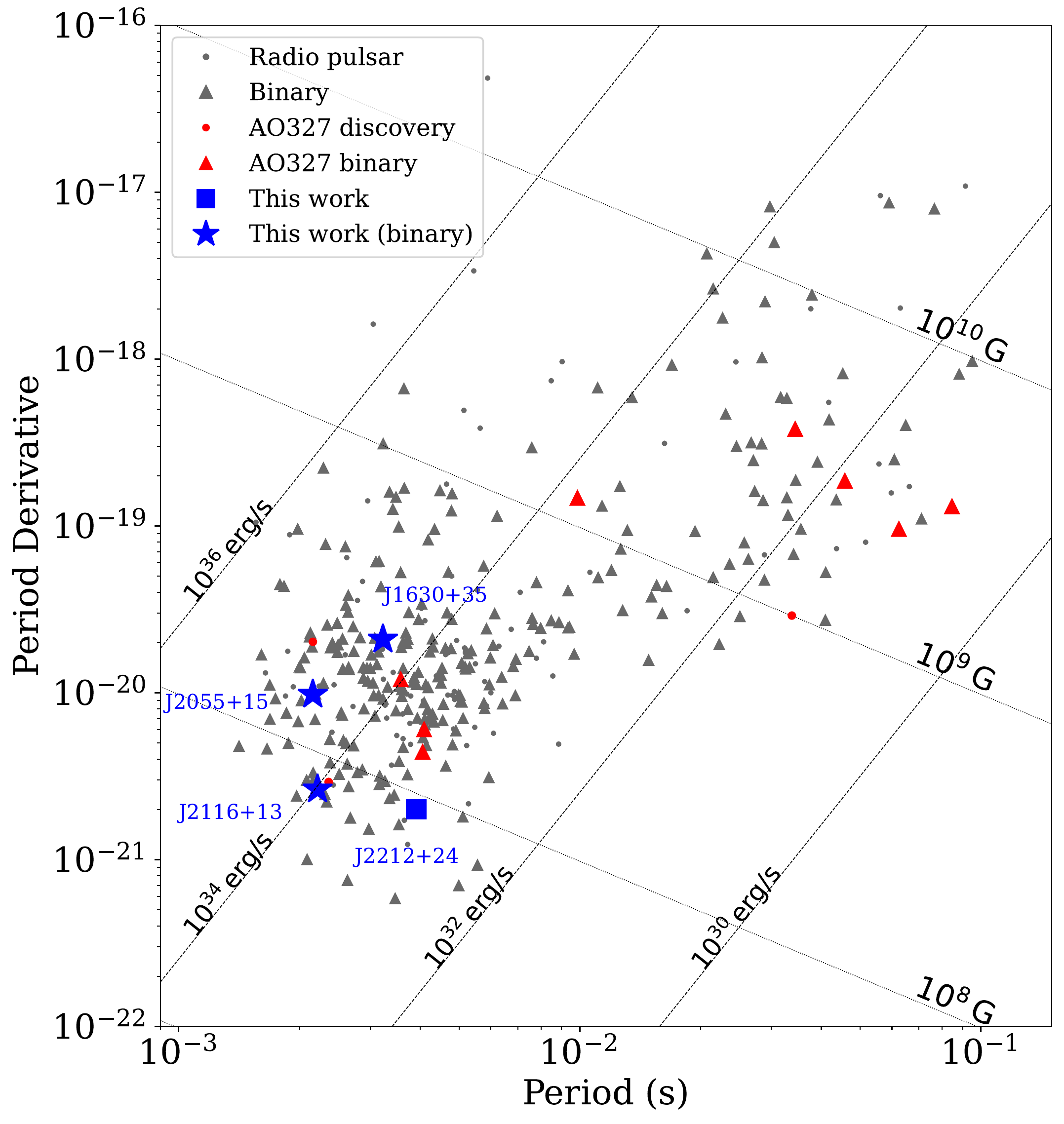}
    \caption{$P-\dot{P}$ diagram of recycled pulsars from the ATNF catalog with $P < 100 \, \rm{ms}$ and $\dot{P} < 10^{-16}$. Pulsars from this work are shown and labeled in blue, and AO327-discovered pulsars are shown in red. Lines of constant minimum surface magnetic field (in Gauss) and spin-down luminosity (in erg s$^{-1}$) are shown.}
    \label{fig:ppdot}
\end{figure}

Figures~\ref{fig:res_date} and~\ref{fig:res_orb}  show the timing residuals (i.e., the difference between observed and model-predicted TOAs of pulses) as a function of arrival time and orbital phase, respectively. No residuals show significant trends, confirming that the timing models presented in this paper accurately describe the pulsars' spin, astrometry, and orbital motion. We also show a $P-\dot{P}$ diagram of our pulsars, as well as all recycled pulsars ($P < 100 \, \rm{ms}$ and $\dot{P} < 10^{-16}$) from the ATNF pulsar catalog \citep{ATNF} in Figure~\ref{fig:ppdot}.

The LOFAR Two-meter Sky Survey second data release (LoTSS-DR2; \citealt{LoTSSDR2}) detected polarized point sources coincident with the positions of PSRs J1630+3550 and J2212+2450. The TULIPP project \citep{LOFAR_TULIPP} commenced a targeted search for pulsars associated with unidentified LoTSS sources in which they independently re-detected both pulsars, and published timing and polarization analyses. We compare our results for each pulsar in their respective sections.

\begin{figure}
    \centering    \includegraphics[width=\linewidth]{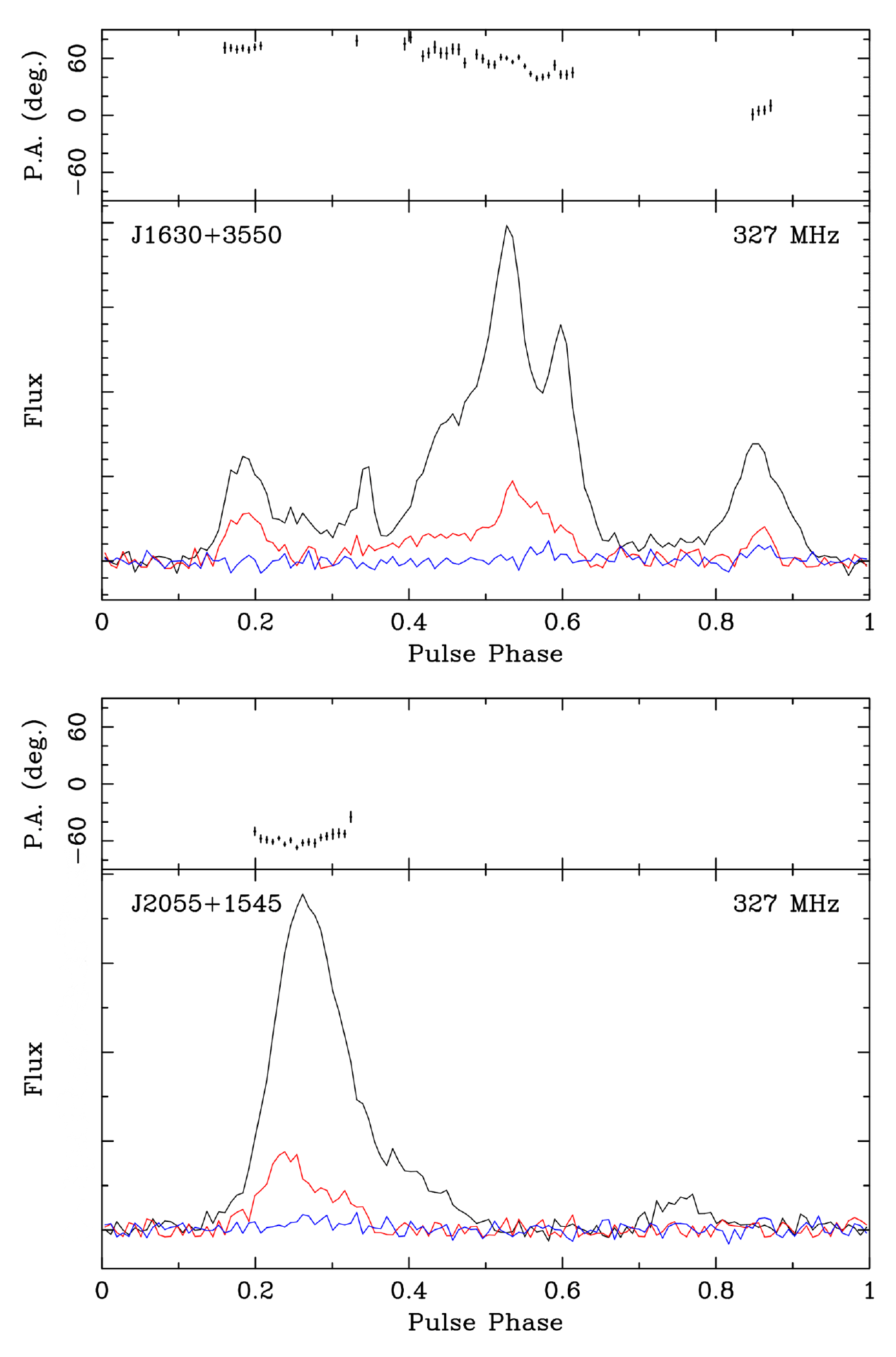}
    \caption{Polarization-calibrated composite profiles for PSRs J1630+3550 (top) and J2055+1545 (bottom), with polarization position angle. The pulse profile for J1630 is centered at a pulse phase of 0.5, with the best-fit RM of 8.72 rad m$^{-2}$ applied. The main component of the profile for J2055 is centered at a pulse phase of 0.25, with the interpulse visible at a phase of 0.75. For J2055, the best-fit RM of $-65.3$ rad m$^{-2}$ has been applied. The $y$-axes are in arbitrary units. Black lines represent total intensity (Stokes \textit{I}), red lines represent linear polarization (Stokes \textit{L}), and blue lines represent circular polarization (Stokes \textit{V}).}
    \label{fig:16302055pol}
\end{figure}

\subsection{PSR J1630+3550}
PSR J1630+3550 (hereafter J1630) is an MSP in a 7.6 hr binary orbit. Assuming a pulsar mass of 1.4 ${M_\odot}$, an orbital inclination angle of $i = 90^{\circ}$ would imply a minimum companion mass of 0.0098 ${M_\odot}$ and a median mass (i.e., $i = 60^{\circ}$) of 0.0113 ${M_\odot}$. The low mass of the companion and the tight orbit indicate that this is a black widow pulsar. In many spider systems, the pulsar's emission is eclipsed by intrabinary material, especially around superior conjunction at an orbital phase of 0.25. We do not observe any eclipses at either 327 or 650 MHz; however, the observations tend to yield lower S/N detections at orbital phases of 0.1-0.3 (see the top panel of Figure~\ref{fig:res_orb}). We do not detect any signature in the residuals due to Shapiro delay  (the very low companion mass precludes this even for high inclinations), and attempting to fit for the delay does not result in a better timing solution. Our best timing solution is displayed in Table~\ref{tab:ELL_solns}. The lack of full eclipses implies that the companion never directly occults the pulsar, and that we are viewing the system relatively face-on.

J1630 was independently re-detected by the LoTSS as a 33\% linearly polarized point source with an RM of 8.65 \rmunits \citep{LoTSSDR2}. Its flux at 144 MHz shows it is a steep spectrum source, to which low-frequency surveys like AO327 and LoTSS are especially sensitive. Our polarization analysis reveals similar properties, with a rotation measure of $8.72 \pm 0.14$ \rmunits. This RM was applied to the polarization-calibrated profile shown in the top panel of Figure~\ref{fig:16302055pol}. Our polarization analysis (and that of \citealt{LOFAR_TULIPP}) reveals a slow variation of the position angle of the linear polarization across the broad central component of the radio profile; this implies that our line of sight traverses the edge of the pulsar's emission beam, rather than cutting through the center.

The 2 yr timing solution for J1630 published by \citet{LOFAR_TULIPP} yielded spin and orbital parameters similar to our solution, and their timing residuals as a function of the orbital phase reveal that at 144 MHz, the system does not exhibit any total eclipses. There is a noticeable increase in the TOA errors in their solution at orbital phases 0.2--0.4; the signal appears to be weakened by intrabinary material near the companion, but the inclination angle of the orbit is likely low enough to preclude total eclipses of the emission.

Further evidence for a possible low inclination angle comes from the broad radio profile, which suggests that our line of sight is closely aligned with the spin axis of the pulsar. Since binary MSPs were spun up by material orbiting in the same plane as the companion star (from where it originated), the angular momentum of the MSP is necessarily aligned with the orbital angular momentum \citep{2023pbse.book.....T}. This implies that, in this system, the orbital angular momentum is also closely aligned with the line of sight, which would again mean a low orbital inclination angle. Furthermore, as mentioned above, the polarization position angle changes slowly across the profile, implying that the distance between the magnetic axis and the line of sight varies slowly with the spin phase. This is consistent with a close alignment between our line of sight and the spin (and thus orbital) axes. These latter arguments depend strongly on the unknown emission geometry of J1630, but lend credence to the overall conclusion that it is likely viewed at a lower inclination angle.

While our solution implies a large transverse velocity on the order of 200 to 300 km s$^{-1}$, the uncertainties on these measurements are large, and the velocities calculated using the NE2001 and YMW16 \citep{ymw17} electron density models are both consistent with transverse velocities as low as 100 km s$^{-1}$. If the velocity of J1630 is as high as the solution suggests, it would still be within the observed range for black widow pulsars (e.g., PSRs J1641+1809 and J2052+1219; \citealt{lsk+18,drc+21}). \citet{dcl+16} found an average 2D transverse velocity of $56 \pm 3$ km s$^{-1}$ for binary pulsars with distances determined through parallax measurements, and an average velocity of $113 \pm 20$ km s$^{-1}$ for binary pulsars with distances estimated using electron density models. In total, the large uncertainties in the proper motion and distance estimates may result in the true proper motion of this system being lower than our current estimate, and closer to the average proper motion for binary pulsars.

\begin{figure}
    \centering
    \includegraphics[width=\linewidth]{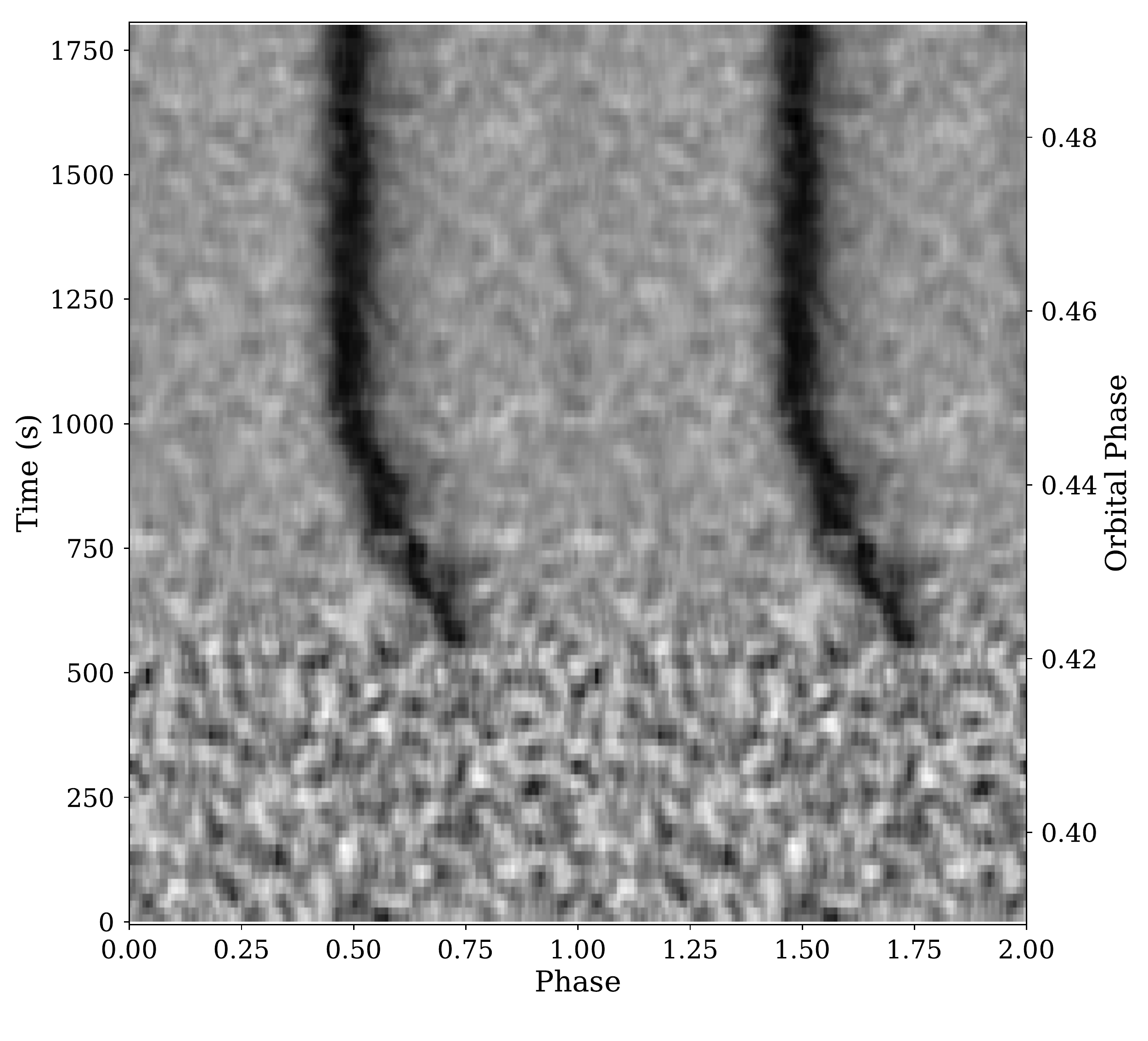}
    \caption{Intensity vs. pulse phase and observation time for J2055 exiting its radio eclipse on MJD 57817, taken with Arecibo at 327 MHz. The pulsed radio emission is delayed from orbital phases 0.42--0.45. Two cycles of the pulse phase are shown.}
    \label{fig:2055eclipse}
\end{figure}

\subsection{PSR J2055+1545}
\label{sec:2055timing}
PSR J2055+1545 (J2055) is an MSP in a circular, 4.8 hr orbit around a companion with a median mass of 0.29 ${M_\odot}$. We observe an eclipse at 327 MHz that spans 36\% of its orbit, which, along with the companion mass, indicates that J2055 is a redback system. In order to unambiguously solve for the orbital evolution of J2055, we used the \textsc{dracula}\footnote{\url{https://github.com/pfreire163/Dracula}} package to build up the timing solution \citep{DRACULA}. We also used the \texttt{spider\_twister}\footnote{\url{https://github.com/alex88ridolfi/SPIDER\_TWISTER}} code to detect the pulsar in several observations \citep{spidertwister}; this was necessary due to the system's high degree of orbital variability.

Our solution spans over 3 yr of Arecibo data and uses the BTX timing model. While we detected this pulsar with the GMRT, we were unable to obtain enough high S/N observations of J2055 to reliably connect the two data sets, and the timing solution presented in this paper was made using only Arecibo observations. Future studies will be able to connect a longer-term solution using the full GMRT data set. Eight orbital frequency derivatives are required to attain well-fitted residuals, which is not uncommon for spider systems due to their orbital variability (e.g. \citealt{drc+21}). The full timing solution is given in Table~\ref{tab:noELL_solns}. 

The radio emission is eclipsed from orbital phases 0.06-0.42, with an orbital phase of 0.25 corresponding to superior conjunction. We also observe delayed pulse arrival during the eclipse egress as shown in Figure~\ref{fig:2055eclipse}, encompassing $\sim3\%$ of the orbit. When the pulsar enters the eclipse we do not observe any pulse delays. Of the GMRT observations on hand, none cover the eclipsing part of the orbit, so we cannot comment on any eclipsing behavior at 650 MHz. In addition to the main eclipse, J2055 exhibits small, short eclipses of its signal throughout its orbit, likely due to stripped gas from the stellar companion moving throughout the system. Previous studies of redbacks (e.g., \citealt{asr+09,ssa16}) also observe short, sporadic eclipses at many points throughout the orbit.

We can place limits on the size of the eclipsing region, given the projected semimajor axis and the length of the eclipses at 327 MHz. Assuming a pulsar mass of 1.4 ${M_\odot}$ and an orbital inclination of 90$^{\circ}$, the orbital separation between the pulsar and companion $a = xc(1+q) = 1.77 \, R_\odot$, with the projected semi-major axis $x$, the speed of light $c$, and the ratio between the pulsar and companion masses $q = m_p / m_c$. The radius of the eclipsing region must then be $R_{\textrm{ec}} > 1.6 \, R_\odot$ in order to obscure the pulsar emission for roughly $36\%$ of the orbit. Comparing this to the Roche lobe of the companion using the relation in \citet{eggleton83}, the radius of the eclipsing region must be at least a factor of 4 larger than the Roche lobe, $R_L = 0.43 \, R_\odot$. These results do not change substantially at lower inclination angles: at an inclination of 60$^{\circ}$, the Roche lobe is still a factor of 4 smaller than the estimated radius of the eclipsing region. Much of the eclipsing material must therefore be gravitationally unbound to the companion, a common feature of redback systems (e.g., \citealt{cls+13,krb+20}).

The emission is at most 25\% linearly polarized in the main component of the profile, with no detected circular polarization. There is also a very weak interpulse, for which no polarization was detected. We measured an RM of $-65.3 \pm 0.16$ \rmunits at 327 MHz, and applied this RM to the polarization-calibrated profile shown in Figure~\ref{fig:16302055pol}.

\subsection{PSR J2116+1345}
PSR J2116+1345 (J2116) is an MSP in a 4.8 days, low-eccentricity ($e \sim 10^{-5}$) orbit with a WD companion with a median mass of 0.22 ${M_\odot}$. We see no evidence of total eclipses at either observing frequency, but the long orbital period precluded us from easily obtaining full orbital coverage.

We obtained an initial timing solution using the combined Arecibo and GMRT radio TOAs, also using the \textsc{dracula} package. The timing-derived position is within the error ellipse of a Fermi gamma-ray source. We detected gamma-ray pulsations using our radio ephemeris and then used the Fermi data to refine the timing solution  and extend it from 2008 August to the end of 2022 (see Section~\ref{sec:fermi}). 
The Fermi solution is given in Table~\ref{tab:ELL_solns}. 

Our polarization analyses at 327, 430, and 1380 MHz all reveal linear polarization in the first component of the profile, and less significant linear polarization in the second component. At 327 MHz, \texttt{rmfit} seemed to equally favor RMs of $\pm 83.1$ \rmunits. At 430 MHz only a peak in the linearly polarized flux at 79.5 \rmunits appeared, and at 1.4 GHz only a peak at $-$89.4 \rmunits appeared. When an RM of 79.5 \rmunits is applied to the 1.4 GHz profile, no linear polarization is detected above $8\%$ of the total intensity, as opposed to over $30\%$ when the best-fit RM of $-$89.4 \rmunits is applied. Similarly, when an RM of $-$89.4 \rmunits is applied to the 430 MHz profile, no linear polarization is detected above $9\%$ of the total intensity, as opposed to $70\%$ when the best-fit RM of 79.5 \rmunits is applied.

\begin{figure*}
    \centering
    \includegraphics[width=\linewidth]{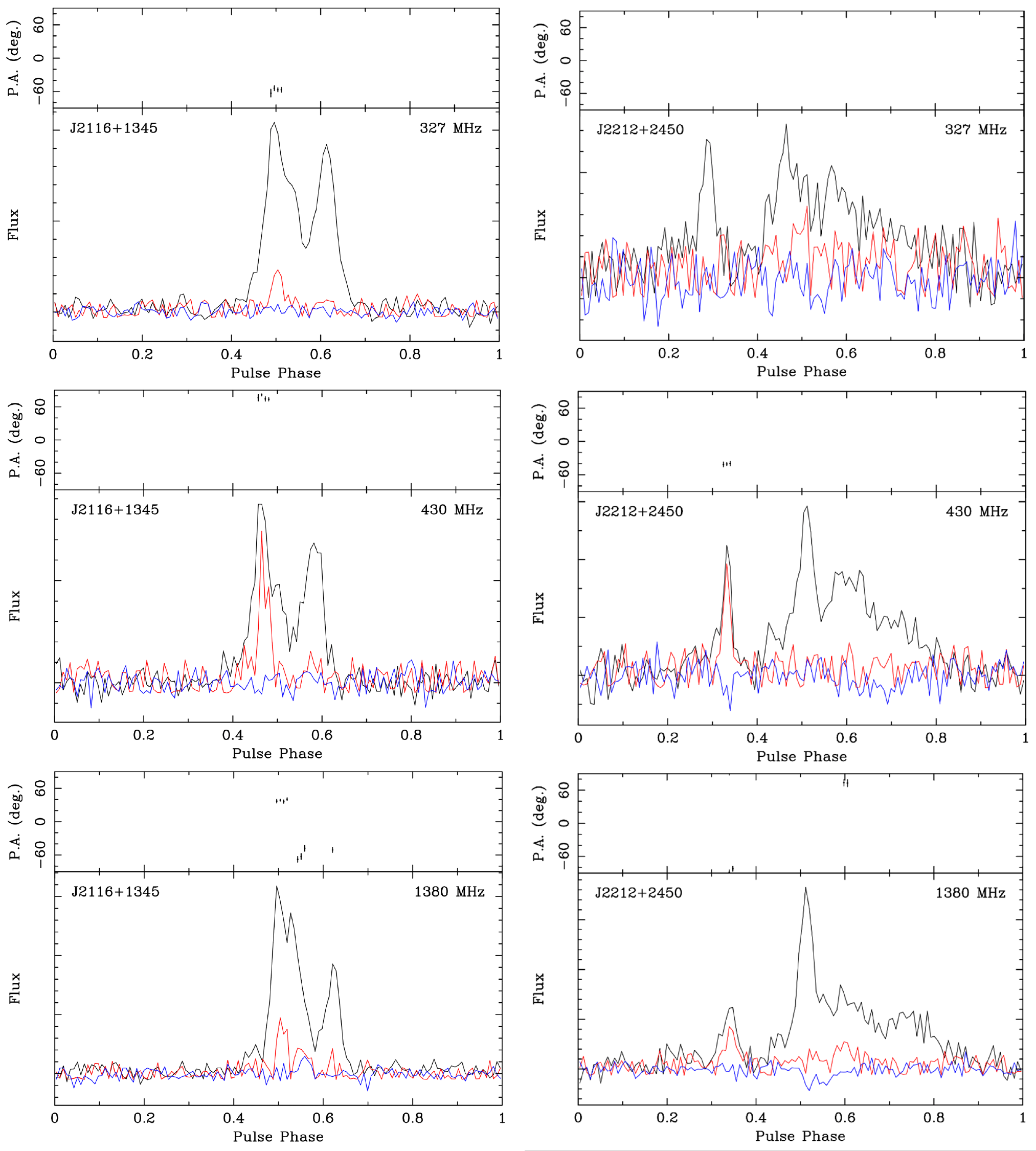}
    \caption{Polarization-calibrated composite profiles for PSRs J2116+1345 (left) and J2212+2450 (right), with polarization position angle, at 327 MHz (top), 430 MHz (middle), and 1380 MHz (bottom). The pulse profiles are centered at a pulse phase of 0.5. The $y$-axes are in arbitrary units. Black lines represent total intensity (Stokes \textit{I}), red lines represent linear polarization (Stokes \textit{L}), and blue lines represent circular polarization (Stokes \textit{V}). The best-fit RMs have been applied to each profile as described in Section~\ref{sec:timing}.}
    \label{fig:2116and2212pol}
\end{figure*}

We chose to adopt the best-fit RMs returned by \texttt{rmfit} at each frequency, applying the positive RM to the 327 MHz data since it yielded a slightly higher detection of linear polarization. Multi-frequency studies of pulsar RMs do not find evidence of RM variation with observing frequency \citep{sbg+19}, and we obtained only one short observation of this pulsar at 430 and 1380 MHz, so we do not draw any conclusions about the true RM of this pulsar or its variation with frequency. At 327, 430, and 1380 MHz, we measure RMs of $83.16 \pm 0.42$, $79.5 \pm 1.7$, and $-89.4 \pm 4.5$ \rmunits, respectively. We adopt the 1$\sigma$ errors on these measurements from \texttt{rmfit}. The polarization-calibrated profiles in Figure~\ref{fig:2116and2212pol} were made after applying these RMs. 

\subsection{PSR J2212+2450}
PSR J2212+2450 (J2212) is the only isolated MSP in this sample, with a rotation period of 3.9 ms and a dispersion measure of 25.2 \dmunits. We have measured a total proper motion of $24 \pm 11$ mas yr$^{-1}$, which implies a total transverse velocity of $180 \pm 100 $ km s$^{-1}$ using the DM distance of 1.59 kpc, calculated from the NE2001 electron density model with an estimated distance uncertainty of 30\% \citep{ne2001}. The full timing solution is given in Table~\ref{tab:noELL_solns}.

J2212 was independently re-detected by the LoTSS as a 6\% circularly polarized point source \citep{LoTSSDR2}, yet no linearly or circularly polarized emission was detected from the TULIPP follow-up observations due to low S/N. However, they were able to detect the pulsar in follow-up observations and confirm its basic spin parameters, which are in agreement with our timing solution \citep{LOFAR_TULIPP}. 

Figure~\ref{fig:2116and2212pol} shows the polarization-calibrated profiles at each observing frequency for J2212. Our polarization analyses were also complicated by the weakness of the pulsar signal; at 327 MHz, we were unable to unambiguously detect any polarization, and no RM from $-$200 to 200 \rmunits was favored by \texttt{rmfit} or applied to the final profile. At 430 MHz and 1.4 GHz, we find strong linear polarization in the narrower component of the profile, but no polarization in the broader component. The best RMs returned by \texttt{rmfit} varied greatly, with peak values of $30.4 \pm 4.7$ and $-44.9 \pm 4.2$ \rmunits for the 430 MHz and 1.4 GHz observations, respectively, even though the observations were taken on the same epoch. The error bars are the 1$\sigma$ errors on the RM fit performed by \texttt{rmfit} with position angle refinement; for both this pulsar and J2116, the true errors on these measurements are likely much larger due to the lack of strongly polarized emission from the pulsar. As with J2116, we only obtained one observation at each frequency with relatively low S/N detections, so we draw no conclusions about the true RM of this source and simply apply the best RM to each profile.

\begin{deluxetable*}{ccccccccccc}
\label{tab:Pdots}
\tablecaption{Corrections to observed period derivatives}
\tablehead{\colhead{PSR} &\colhead{$\mu_T$} & \colhead{$v_t$} & \colhead{$D_{DM}$} & \colhead{Observed $\dot{P}$} & \colhead{$\dot{P}_{\rm{shk}}$} & \colhead{$\dot{P}_{\rm{gal}}$} & \colhead{$\dot{P}_{\rm{int}}$} & \colhead{$\dot{E}$} & \colhead{$B_{\rm{surf}}$} & \colhead{$\tau$} \\
\colhead{} & \colhead{(mas yr$^{-1}$)} & \colhead{(km s$^{-1}$)} & \colhead{(kpc)} & \colhead{($10^{-21}$)} & \colhead{($10^{-21}$)} & \colhead{($10^{-21}$)} & \colhead{($10^{-21}$)} & \colhead{($10^{34}$ erg s$^{-1}$)} & \colhead{($10^8$ G)} & \colhead{(Gyr)}}
\startdata
\multirow{2}*{J1630+3550} &  \multirow{2}*{40(24)} 
 & 210(140) & 1.1 & \multirow{2}*{22.55(5)} & 14(12) & $-$0.5(5) & 9(12) & \nodata & \nodata & \nodata \\
& & 300(200) & 1.57 & & 19(18) & $-$0.59(6) & 4(18) & \nodata & \nodata & \nodata \\
\multirow{2}*{J2055+1545}  & \multirow{2}*{4.0(7)} & 46(16) & 2.44 & \multirow{2}*{9.772(8)} & 0.20(8) & $-$0.5(1) & 10.0(1) & 3.9 & 1.5 & 3.4 \\
 & & 69(24) & 3.66 & & 0.3(1) & $-$0.8(2) & 10.2(2) & 4.0 & 1.5 & 3.3 \\
\multirow{2}*{J2116+1345}  & \multirow{2}*{11(3)} & 100(40) & 1.85 & \multirow{2}*{2.643(7)} & 1.2(6) & $-$0.41(8) & 1.8(6) & 0.65 & 0.64 & 19 \\
 & & 140(60) & 2.73 & & 1.8(9) & $-$0.6(1) & 1.5(9) & 0.54 & 0.58 & 23 \\
\multirow{2}*{J2212+2450} & \multirow{2}*{24(11)} & 180(100) & 1.59 & \multirow{2}*{2.0(6)} & 9(6) & $-$0.7(1) & $-$6(6) & \nodata & \nodata & \nodata \\
 & & 240(130) & 2.09 & & 12(8) & $-$0.9(2) & $-$9(8) & \nodata & \nodata & \nodata
\enddata
\tablecomments{Proper motions, transverse velocities, distance estimates, and spin-down corrections for each pulsar using the NE2001 (top) and YMW16 (bottom) electron density models, assuming $30\%$ error on the distance estimates. Using the calculated intrinsic spin-down rate, we recalculate the spin-down luminosity, surface dipole magnetic field, and characteristic age for PSRs J2055+1545 and J2116+1345 (see Section~\ref{sec:pdotcorr}). The 1$\sigma$ error bars are denoted with parenthesis on the last significant digit.}
\end{deluxetable*}

\subsection{Intrinsic Period Derivative Corrections}
\label{sec:pdotcorr}
The observed spin-down rate $\dot{P}_{\rm{obs}}$ of a pulsar can deviate from its intrinsic spin-down rate $\dot{P}_{\rm{int}}$, due to the pulsar's transverse motion as well as Galactic acceleration. Using distance and proper motion measurements, we can calculate the kinematic corrections to the pulsar's spin-down rate using 
\begin{equation}
    \dot{P}_{\rm{int}} = \dot{P}_{\rm{obs}} - \dot{P}_{\rm{Shk}} - \dot{P}_{\rm{gal}}.
\end{equation}

$\dot{P}_{\rm{Shk}}$ represents the correction from the Shklovskii effect \citep{Shk}, an apparent spin-down due to the pulsar's motion relative to the solar system barycenter, calculated using
\begin{equation}
    \dot{P}_{\rm{Shk}} = \frac{P}{c} \mu_T^2 d,
\end{equation}

where $P$ is the pulsar's rotational period, $\mu_T$ the total transverse proper motion, and $d$ the distance to the pulsar. We calculate $\dot{P}_{\rm{gal}}$, the spin-down correction from Galactic acceleration, using the same procedure as described in \citet{gfg+21}, which uses an approximation for the vertical component of Galactic acceleration \citep{hf04}. The approximation is valid for pulsars within a Galactic height of $\sim \pm 1.5$ kpc, which is true for all the pulsars here. We use the latest values of the distance to the center of the Galaxy ($R_0 = 8.275 \pm 0.034 $ kpc; \citealt{rb20}) as well as the circular velocity of the Sun ($\Theta_0 = 240.5 \pm 4.1$ km s$^{-1}$, \citealt{Gravity}). 

Table~\ref{tab:Pdots} lists the proper motions, transverse velocities, and distance estimates for each pulsar using both the NE2001 and YMW16 electron density models. Using these values, we calculate the spin-down corrections from the Shklovskii effect and Galactic acceleration, and subtract these from the observed spin-down rate to find the pulsar's intrinsic spin-down rate, $\dot{P}_{\rm{int}}$. Using this new value, we recalculate the spin-down luminosity $\dot{E}$, surface magnetic field $B_{\rm{surf}}$, and characteristic age $\tau$. We present the calculated intrinsic spin-down rates for each pulsar, but the large uncertainties in the proper motion measurements for PSRs J1630 and J2212 lead to physically insignificant estimates of $\dot{P}_{\rm{int}}$, so we do not recalculate the derived parameters for those pulsars.

The imprecise proper motion measurements for PSRs J1630 and J2212 lead to large uncertainties in the estimate of the contribution from the Shklovskii effect; this leads to $\dot{P}_{\rm{int}}$ values smaller than their 1$\sigma$ error bars in 
the case of J1630, and negative $\dot{P}_{\rm{int}}$ values in the case of J2212.  Fractional errors of up to $50\%$ are not uncommon for distances estimated using the NE2001 and YMW16 electron density models \citep{dgb+19}. Due to the large distance uncertainties in these models, independent distance measurements of these pulsars could be useful, which can be obtained from parallax measurements from timing or from very long baseline interferometry experiments. Longer-term timing solutions could also constrain the proper motions more significantly and measure the intrinsic spin-down rates of these pulsars.

\section{MultiWavelength Follow-up}
\label{sec:counterparts}

\subsection{Gamma-ray Counterparts}
\label{sec:fermi}
\begin{figure}
    \centering
    
    \includegraphics[width=\linewidth]{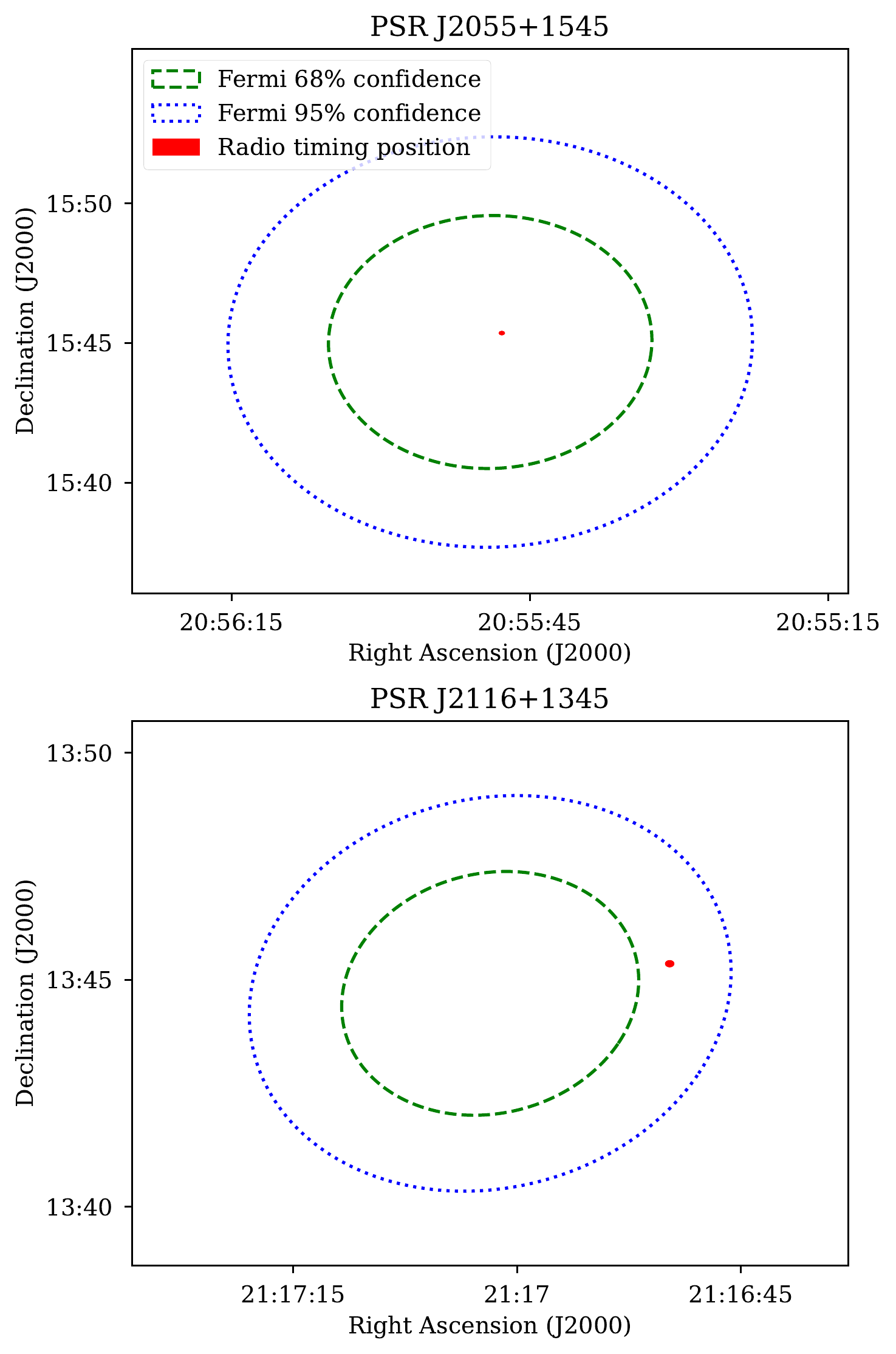}
    \caption{The radio timing positions for PSRs J2055+1545 (top) and J2116+1345 (bottom) are represented by red ellipses, plotted against the 68\% (green dashed line) and 95\% (blue dotted line) confidence ellipses for their potentially coincident Fermi sources. The ellipses representing the radio timing positions are roughly 2000 times the size of the 1$\sigma$ uncertainties in the sky positions derived from pulsar timing.}
    \label{fig:ellipses}
\end{figure}

We searched for gamma-ray counterparts for each pulsar in the 12 yr iteration (4FGL DR3; \citealt{4FGLDR3}) of the Fermi-LAT point source catalog (4FGL; \citealt{Fermi4SC}).  Both PSRs J2055+1545 and J2116+1345 fell within the Fermi-LAT 95\% error regions of Fermi sources; Figure~\ref{fig:ellipses} shows the 68\% and 95\% error ellipses compared with the sky positions derived through the radio timing solutions. 

PSR J2116+1345 is coincident with 4FGL J2117.0+1344, a gamma-ray point source with a 0.1-100 GeV flux of 2.9(4)$\times 10^{-12}$ erg cm$^{-2}$ s$^{-1}$. This flux corresponds to gamma-ray luminosities of $L_{\gamma}$ = 0.12 and 0.26 $\times 10^{34}$ erg s$^{-1}$ using the NE2001 and YMW16 DM-derived distances, respectively. Comparing this to the pulsar's intrinsic spin-down powers calculated in Table~\ref{tab:Pdots} suggests a gamma-ray efficiency of  $L_{\gamma} / \dot{E} \approx 35 \pm 15 \%$, a reasonable value for gamma-ray MSPs \citep{Fermi2PC}. The Fermi analysis pipeline uses the positions of the Fermi point source and any potential counterparts to calculate an association probability using Bayesian statistics \citep{Fermi4SC}; this Fermi source and PSR J2116+1345 have an association probability of 62\%. 

We performed a search for gamma-ray pulsations by weighting each photon according to its probability of being associated with the source \citep{mk11}. With our best radio timing solution, we used the \texttt{fermi} plugin to \textsc{tempo2} to search around the pulsar's timing parameters to find the values which maximized the significance of gamma-ray pulsations using the weighted $H$-test statistic \citep{db10}. Applying this method to the Fermi data ranging from the start of the mission until the end of 2022, we obtain a long-term timing solution for J2116 which we present in Table~\ref{tab:ELL_solns}.

\begin{figure}
    \centering
    
    \includegraphics[width=\linewidth]{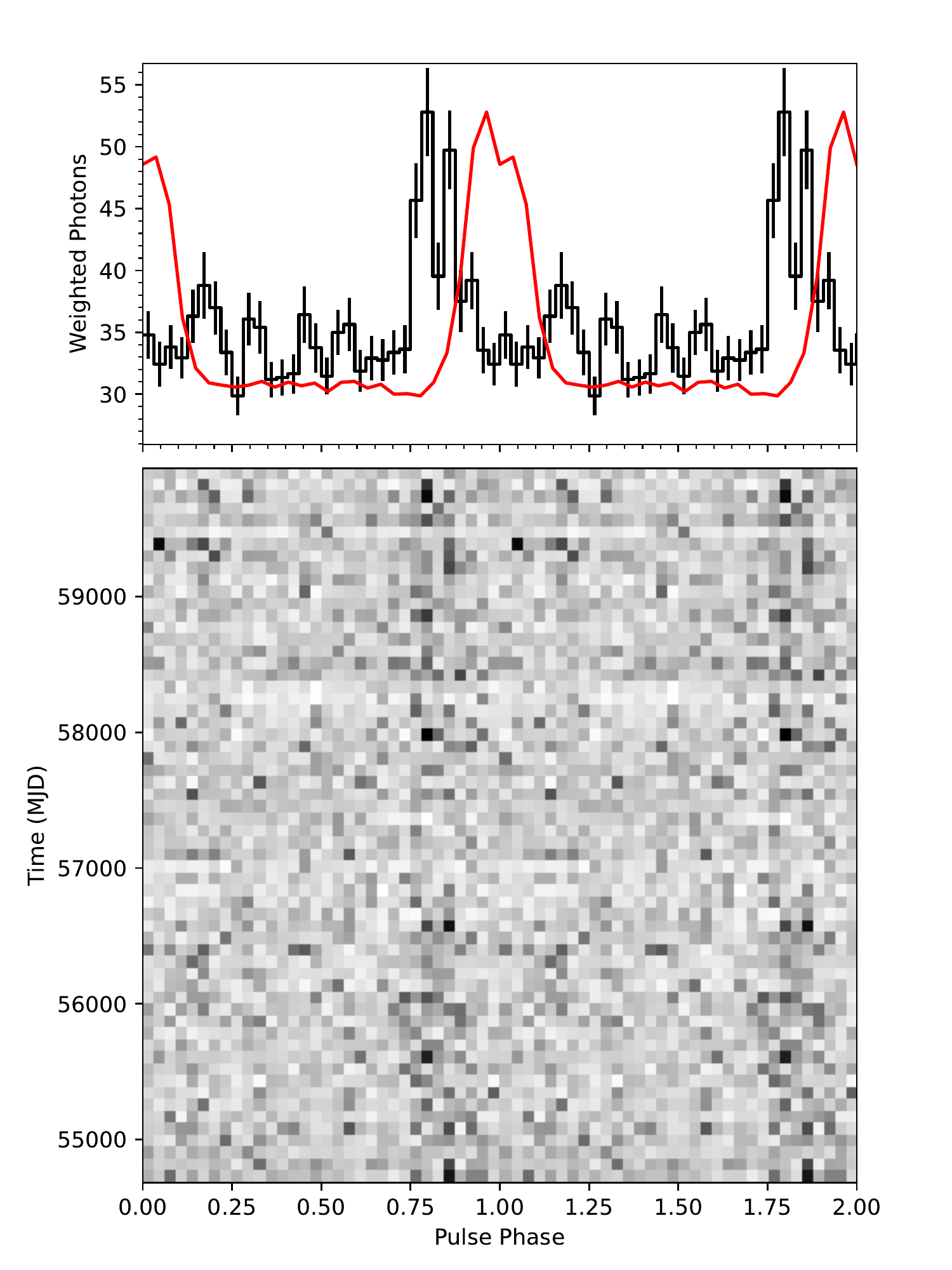}
    \caption{Top panel: phase-aligned integrated Fermi gamma-ray pulse profile (black) and 327~MHz Arecibo profile (red). Bottom panel: phase-time diagram for the gamma-ray pulsations from PSR J2116+1345. Two cycles of the pulse phase are shown.}
    \label{fig:fermi}
\end{figure}

Figure~\ref{fig:fermi} shows the phase histogram of gamma-ray pulsations and the integrated gamma-ray profile made with the Fermi timing solution; the fold yields a significance of 11.4$\sigma$. The 327 MHz integrated pulse profile is phase-aligned and overplotted on top of the gamma-ray profile. We observe a main gamma-ray peak with a complex morphology, the center of which trails the peak of the radio profile by a phase of roughly 0.2. Misaligned profiles such as this are often explained by \textit{outer gap} or \textit{slot gap} models, in which the radio emission emerges from particle acceleration along open magnetic field lines above the pulsar's polar caps, while gamma-ray radiation is emitted at higher altitudes in the pulsar magnetosphere in the region surrounding the open field lines (see \citealt{akh22} for a recent review).

PSR J2055+1545 falls within the 68\% error ellipse of the Fermi source 4FGL J2055.8+1545 (see Figure~\ref{fig:ellipses}), corresponding to a Bayesian association probability of 99\%.  Despite a thorough search for gamma-ray pulsations using our best radio-derived timing solution, no pulsations were detected above a significance level of 2.5$\sigma$. The orbital variability of this system makes it difficult to reliably search for phase-aligned gamma-ray pulsations outside of the range of the radio timing solution. Given the high association probability, continued radio timing of this source may yield better results.

\begin{deluxetable}{cc}
\label{tab:optical}
\tabletypesize{\footnotesize}
\tablecaption{The optical magnitudes measured by the three optical surveys which detected J2055+1545. }
\tablehead{}
\startdata
\bf{GAIA DR3}\tablenotemark{a} & \citet{GaiaDR3} \\
Source ID & 1763537692275731328 \\
Red (${G_{\mathrm{RP}}}$) magnitude & $20.07 \pm 0.09$ \\
Green ($G$) magnitude & $20.50 \pm 0.01$ \\
Blue (${G_{\mathrm{BP}}}$) magnitude & $20.8 \pm 0.1$ \\
\hline
\bf{Pan-STARRS} & \citet{PanSTARRS} \\
Source ID & 126903139493667392 \\
\textit{g} magnitude & $21.013 \pm 0.084$ \\
\textit{r} magnitude & $20.393 \pm 0.036$ \\
\textit{i} magnitude & $20.293 \pm 0.036$ \\
\textit{z} magnitude & $20.272 \pm 0.018$ \\
\textit{y} magnitude & $20.00 \pm 0.13$ \\
\hline
\bf{SDSS DR17} & \citet{SDSSDR17} \\
Source ID & 1237666428821307499 \\
\textit{u} magnitude & $22.81 \pm 0.31$ \\
\textit{g} magnitude & $21.040 \pm 0.031$ \\
\textit{r} magnitude & $20.370 \pm 0.027$ \\
\textit{i} magnitude & $20.184 \pm 0.035$ \\
\textit{z} magnitude & $20.26 \pm 0.14$ \\
\enddata
\tablenotemark{a}{Taken from the Aladin Lite interface. Reported errors on the magnitudes are taken from the \texttt{phot\_mean\_mag\_error} entries.}
\end{deluxetable}

\subsection{Optical Counterparts}
\label{sec:optical}
We used the Aladin Lite \citep{Aladin} interface to search for optical counterparts within 5 arcseconds of the most up-to-date position in each timing solution. The only pulsar with an optical counterpart within this error ellipse was PSR J2055+1545, which is coincident with a magnitude $G=20.5$ source identified in Gaia Data Release 3 \citep{GaiaDR3}. We used the Gaia measurement of the companion's proper motion in the radio timing solution, keeping the values fixed.  At this time, Gaia has not been able to measure a parallax for the companion. 

The Gaia source is offset by the radio timing position by only $0\,\farcs015$. In Gaia DR3, the density of sources with $G$ magnitudes above 20.5 is roughly $2.5 \times 10^{-3}$ arcsec$^{-2}$; this leads to a false association probability of 2 $\times 10^{-6}$. The association between the pulsar and this optical source is extremely convincing, especially given that we use the proper motion measured by Gaia in our phase-connected timing solution.

The counterpart has also been identified by the Sloan Digital Sky Survey (SDSS) Data Release 17 \citep{SDSSDR17} and the Pan-STARRS Data Release 2 \citep{PanSTARRS} optical surveys. Details of the photometric measurements from each survey are given in Table~\ref{tab:optical}. We discuss the properties of this counterpart in the context of the rest of the properties of J2055 in Section~\ref{sec:2055disc}.

\subsection{X-ray Followup}
\label{sec:xray}
We observed the region around PSR J2055+1545 with the Swift X-Ray Telescope for 8600 s on 2022 23 March in the photon counting mode. No X-ray emission was detected from a 15$\arcsec$ region centered on the pulsar's position. Assuming a photon index of 1.5 and an interstellar neutral hydrogen absorption\footnote{Calculated using the HI column density tool located at \url{https://heasarc.gsfc.nasa.gov/cgi-bin/Tools/w3nh/w3nh.pl}} of $6.6 \times 10^{20}~\rm{cm}^{-2}$, a single photon detection in 8600 s would correspond to an upper flux limit of $5.7 \times 10^{-15}~\rm{erg\, cm^{-2}\, s^{-1}}$ in the 0.3-10 keV band. Using DM-derived distances of 2.4 and 3.7 kpc for the NE2001 and YMW16 models, respectively, these correspond to X-ray luminosity upper limits of $3.9 \times 10^{30}$ and $9.3 \times 10^{30}~\rm{erg\, s^{-1}}$. We discuss this result in Section~\ref{sec:2055disc}.

\section{Discussion}
\label{sec:disc}

\begin{figure}
    \centering
    
    \includegraphics[width=\linewidth]{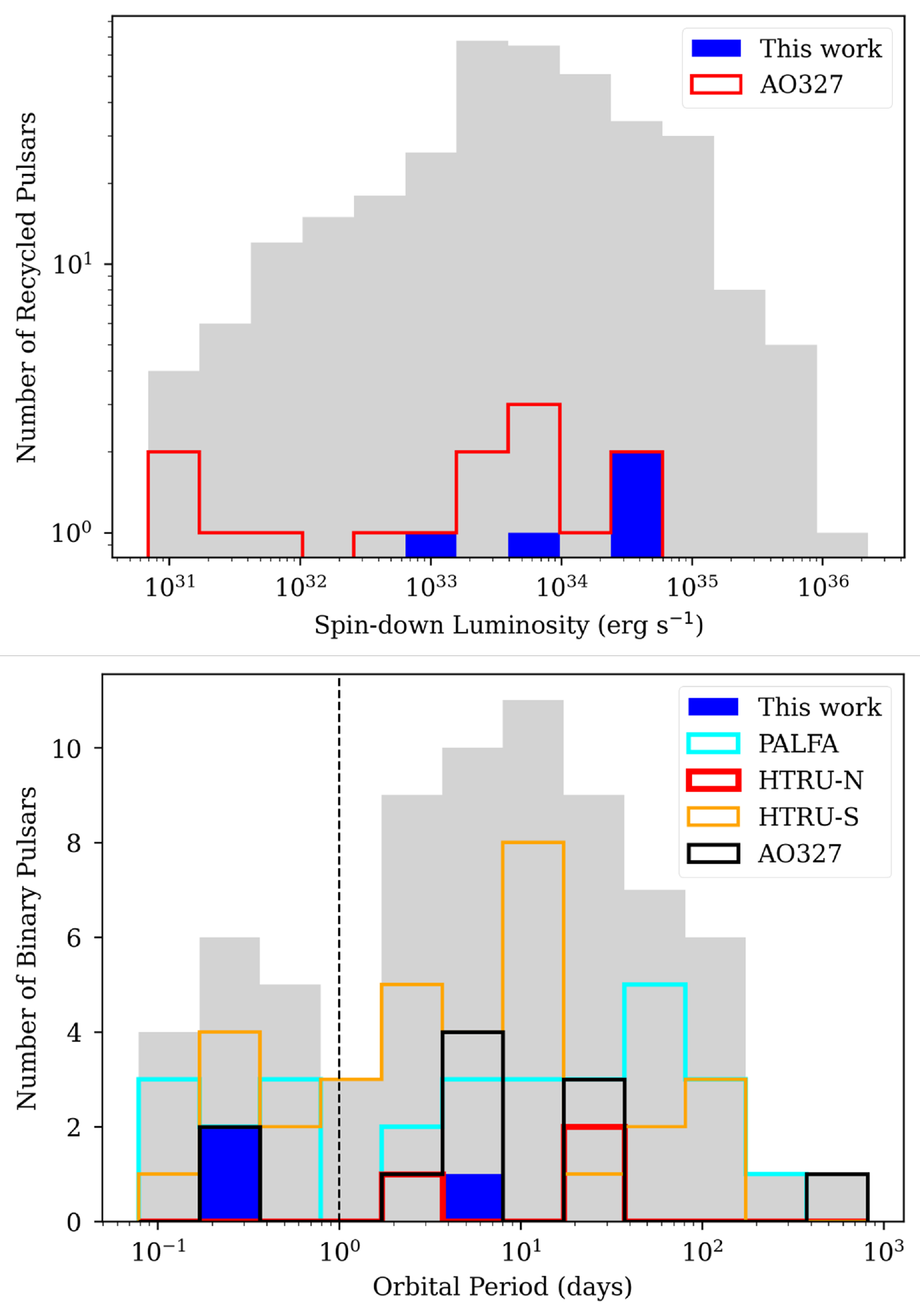}
    \caption{Top panel: in gray, the distribution of spin-down luminosities ($\dot{E}$) for recycled pulsars with $P < 100\, \rm{ms}$ and $\dot{P} < 10^{-17}$, with the distribution for AO327-discovered pulsars outlined in red and the pulsars from this work in blue. Bottom panel: the distribution of orbital periods for binary pulsars discovered by the PALFA (cyan), HTRU-North (red), HTRU-South (orange), and AO327 (black) surveys, with the pulsars from this work in blue. The total distribution from these four surveys is plotted in gray, and the vertical dashed black line represents an orbital period of 1 day.}
    \label{fig:surveycompare}
\end{figure}

\subsection{AO327 Pulsar Population}
At the low observing frequency of AO327, astrophysical signals are more strongly affected by dispersion and scattering by the interstellar medium. For nearby pulsars with low DMs, the signal will be dominated by dispersion rather than scattering, and we can retain sensitivity by searching many trial DMs and minimizing downsampling of the data \citep{dsm+13}. On the other hand, high-DM signals are more strongly broadened by interstellar scattering, which cannot be easily corrected for in the data processing stages. Scattering especially impacts the detectability of fast pulsars with high DMs at low frequencies; according to \citet{bcc+04}, for a DM of 100 \dmunits and central frequency of 327 MHz, the scattering timescale is $\tau_{\rm s} = 1$ ms. 

We expect that of the recycled pulsars ($P < 100 \, \rm ms$) discovered by AO327, they will have relatively low DMs and thus comprise a more local, low-luminosity pulsar population. The top panel of Figure~\ref{fig:surveycompare} compares the distribution of spin-down luminosities ($\dot{E}$) for AO327-discovered recycled pulsars against all recycled pulsars from the ATNF catalog (with $P < 100 \, \rm ms$ and $\dot{P} < 10^{-17}$). It is clear that recycled pulsars discovered by AO327 tend to have a lower spin-down power than the general population.

Given the drift-scan nature of the AO327 survey, a source will remain in the telescope beam for approximately 1 minute as Earth rotates the beam across the sky. The short integration time means that our survey is more sensitive to highly accelerated binaries with short orbital periods, especially if an acceleration search is used. This should imply that the binary systems discovered by AO327 tend to have shorter orbital periods. The bottom panel of Figure~\ref{fig:surveycompare} compares the binary periods of AO327-discovered binaries to those in the ATNF catalog discovered by the PALFA \citep{PALFA}, HTRU-North \citep{HTRUN}, and HTRU-South \citep{HTRUS} surveys. Contrary to the expectation that AO327-discovered binaries would skew towards shorter orbital periods, we see no evidence for this detection bias. In fact, this work presents the first two binaries discovered by AO327 with orbital periods shorter than a day.

\begin{figure*}
    \centering
    \includegraphics[scale=0.55]{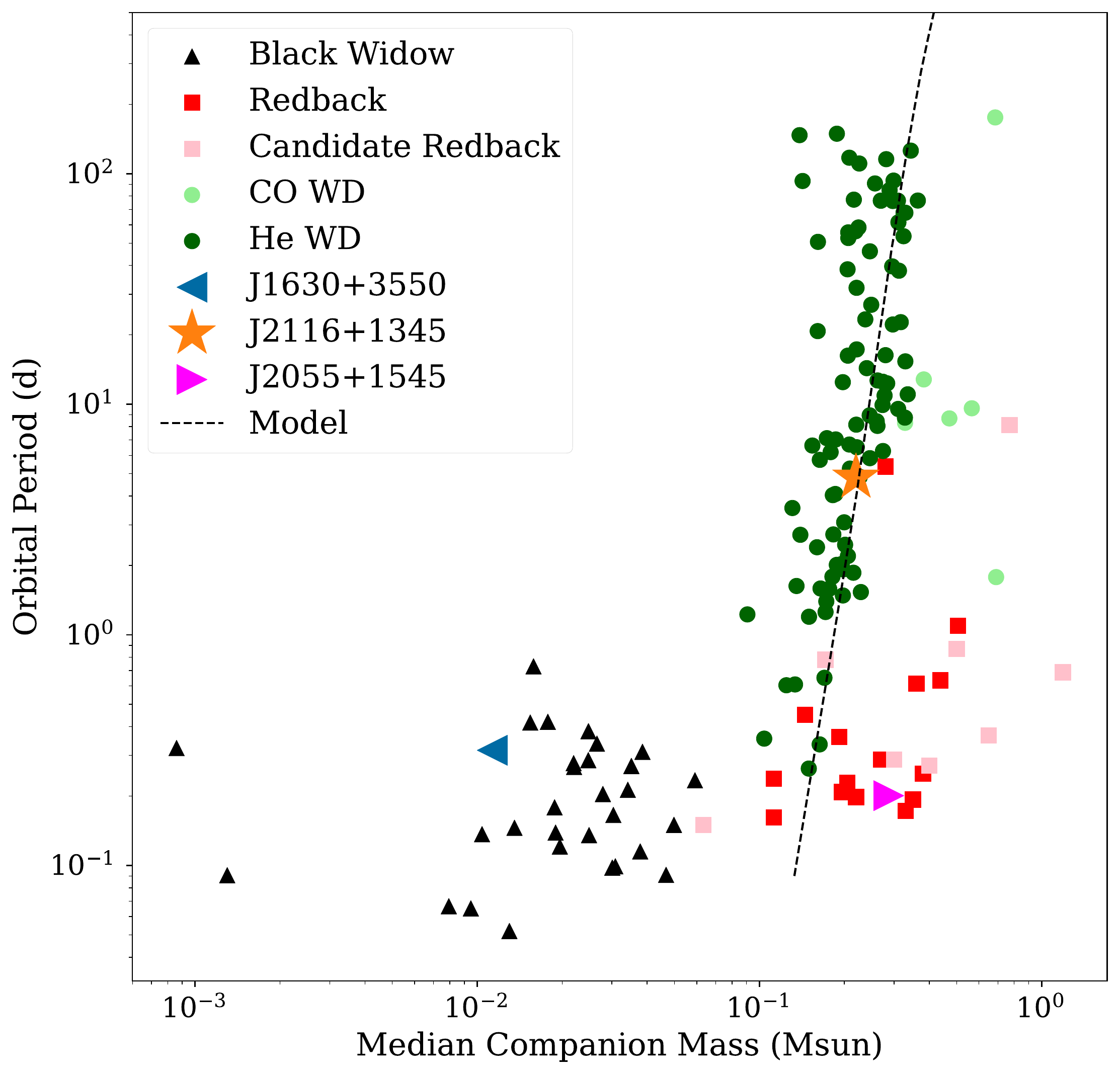}
    \caption{The orbital periods and median companion masses for recycled MSPs with known companion star types. The black dashed line represents the evolutionary models of NS binaries containing He WDs calculated by \citet{ts99}, and the He WDs (dark green circles) are shown to closely follow this relation. Black widows are shown with black triangles, and redbacks are shown with red squares. Figure and caption adapted from \protect{\citet{ssc+19}} and \protect{\citet{rnc+22}}.}
    \label{fig:OPCM}
\end{figure*}

\subsection{Binary Properties}
Figure~\ref{fig:OPCM} shows the orbital period versus companion mass for binary MSPs with known companion star types, plotted against the expected evolutionary track for binaries containing He WDs, as simulated by \citet{ts99}. In particular, PSR J2116+1345 lines up with this model well. The agreement suggests, additionally, that the orbital inclination is near the value of $60^\circ$ used for the median estimate of companion masses. Furthermore, the eccentricity is consistent with the predictions of \cite{esp92}. The non-detection of eclipses is expected for a MSP - He WD system.

PSR J1630+3550 has similar parameters to other black widow systems, with a slightly lower median companion mass than average. As discussed in Section~\ref{sec:timing}, this system may be viewed more face-on at a low inclination angle, meaning that the true companion mass may be higher than as shown in Figure~\ref{fig:OPCM}. Among black widow systems, a correlation between a low mass function and lack of eclipses has been observed by \cite{fck+03}, who proposed that this could be explained if we are seeing the systems with low mass functions at lower inclinations; this would imply that the companions to black widow systems have a range of masses that is significantly narrower than suggested by the range of mass functions.

PSR J2055+1545 exhibits similar properties to the remainder of the redback population, with a relatively short orbital period, broad radio eclipses that last almost half of the orbit and large, unpredictable variations of the orbital period.

\subsection{The Black Widow System J1630+3550}
Despite the relative proximity ($\lesssim 2$ kpc) of J1630 to Earth, no optical, X-ray, or gamma-ray counterparts have been reported. Out of a sample of 37 black widow systems and four candidate systems, 21 have optical photometry measurements, 20 have been observed in X-rays, and 33 have been detected in gamma rays with Fermi \citep{ssc+22}. Many black widow and redback systems have optical counterparts with highly variable orbital light curves due to irradiation of the substellar companion by the pulsar wind. The orbital light curves of many black widows have a maximum when the heated side of the companion faces Earth, and a minimum when it faces away from Earth (e.g., \citealt{drf+19, zkz+19}). At higher energies, X-ray emission in spider systems can arise from a shock between the pulsar wind and the companion's wind or magnetosphere, and most MSPs (spiders or not) appear to be gamma-ray emitters \citep{Fermi2PC}.

The pulsar's spin-down power of $\dot{E} \sim 10^{34} $\, erg s$^{-1}$ is low compared to MSPs with detected nonthermal X-ray pulsations, and the lack of thermal X-ray emission suggests the lack of a significant shock between the pulsar wind and ablated material in the system. Furthermore, the lack of a detected optical counterpart suggests that the companion of J1630 is extremely dim, even with brightened emission due to heating from the pulsar. Of the 30 black widow systems with measured $\dot{E}$, J1630 has a spin-down luminosity close to the median value of $2.1 \times 10^{34} $ erg s$^{-1}$, and its projected semimajor axis of 0.03 lt-s is typical among black widows \citep{ssc+22}. Thus, we conclude that the pulsar wind is of fairly average strength, but there is little interaction between the pulsar emission and the companion's magnetosphere, wind, or ionized material elsewhere in the system. In summary, these findings are consistent with PSR J1630+3550 being a black widow system with a low-mass WD companion, much of which has been ablated away by the pulsar's particle wind.

\subsection{The Redback System J2055+1545}
\label{sec:2055disc}
Using the photometric measurements of the SDSS DR17 given in Table~\ref{tab:optical}, correcting for reddening using \citet{gsf+15}, we convert the $g-r$, $g-i$, and $g-z$ colors into effective temperatures using the equations given by \citet{pam+12}. The SDSS filters give approximate effective temperatures of 5260, 5275, and 5540 K respectively. Furthermore, the relationships between spectral type and color given by \citet{cis+07} indicate that the stellar companion is a late G- or early K- type star. These stars are typically three times as massive as J2055's median companion mass, but due to the irradiation of the star by the pulsar wind, redback companions tend to be hotter and less dense than main-sequence stars with the same mass (e.g., \citealt{lht+14}), so its color and temperature likely do not follow main-sequence relationships. The size of the eclipsing region is substantially larger than the Roche lobe radius of the companion (see Section~\ref{sec:2055timing}), further indicating that the pulsar wind has stripped material from the stellar companion. This irradiation also heats and illuminates the pulsar-facing side of the companion; obtaining orbital light curves and detailed spectroscopy of the stellar companion to J2055 may help constrain the companion mass and stellar type, as well as the system inclination.

Many spider systems have been observed to emit bright, nonthermal X-ray emission, usually attributed to an intrabinary shock between the pulsar wind and stellar wind (e.g., \citealt{vwv20}). Redbacks in particular tend to have brighter, harder X-ray emission, likely due to stronger winds from the larger companions \citep{ssc+22}. Unlike many redbacks, J2055 has no detected X-ray emission: using its upper luminosity limit derived in Section~\ref{sec:xray} implies a 0.3-10 keV luminosity of less than $2.5 \times 10^{-4}~ \dot{E}$. If the distance to the pulsar were underestimated and the true distance was 5 kpc, then the X-ray non-detection would suggest a luminosity of up to $5 \times 10^{-4}~ \dot{E}$. These values are low, but not anomalous, compared to the redback population \citep{ssc+19}. The lack of detectable X-ray emission from J2055 suggests a less significant intrabinary shock, more typical of standard binary MSPs and less typical of \textit{spider} pulsars \citep{rmg+14}. We are unable to place severely constraining limits on the X-ray luminosity of J2055, but deeper observations may reveal X-ray emission.

Despite the colocation of PSR J2055+1545 with the Fermi source 4FGL J2055.8+1545, our inability to detect gamma-ray pulsations leaves us unable to confirm this as a counterpart. There have been 11 discovered MSPs (including J2055) that happen to lie in Fermi error ellipses without detected gamma-ray pulsations \citep{Fermi3PC}. 

The Fermi 4FGL catalog lists a 0.1-100 GeV flux of 2.3(6)$\times 10^{-12}$ erg cm$^{-2}$ s$^{-1}$, for the potentially associated Fermi source 4FGL J2055.8+1545. This flux corresponds to gamma-ray luminosities of 1.6 and 3.7 $\times 10^{33}$ erg s$^{-1}$ using the NE2001 and YMW16 DM-derived distances, respectively. Assuming a $10\%$ efficiency of conversion from spin-down luminosity to gamma-ray luminosity (the median efficiency of 10 redback systems; Figure~11 from \citealt{ssc+19}), we would expect a gamma-ray luminosity on the order of 10$^{33}$ erg s$^{-1}$, lending credence to the possibility of a genuine association. We also note that the flux of the Fermi source is relatively stable over time, typical of gamma-ray pulsars \citep{Fermi2PC}. While the spectral energy distribution is not currently well-constrained, it appears to be pulsar-like: the spectral energy density peaks at roughly 1.4 GeV and falls off at higher energies.

Due to the pulsar-like properties of the Fermi source and the close colocation of the source and the pulsar's timing position, we find convincing evidence that despite the lack of detected gamma-ray pulsations, the Fermi source is associated with J2055. The many orbital frequency derivatives necessary in the timing solution suggest a highly variable orbit, which would complicate the phase alignment outside of the radio timing solution. It is possible that a longer-term timing solution would allow for a detection of gamma-ray pulsations and confirm this association. 

\section{Summary and Conclusion}
\label{sec:conclusion}
In this paper, we present the discovery of four MSPs with the Arecibo 327 MHz Drift-Scan Survey and obtain phase-connected timing solutions using observations with the Arecibo Telescope and the GMRT. Three of these pulsars are in binary systems, consisting of an eclipsing redback, a non-eclipsing black widow, and a NS-WD system. The systems have orbital periods ranging from 4.8 hr to 4.8 days.

The black widow system PSR J1630+3550 shows no evidence of total eclipses at 327 or 650 MHz. The pulsar shows radio emission across its rotation, with four distinct components. Along with the shallow position angle curve, we conclude that the system may be viewed at a low inclination angle, with a low-mass WD companion. We do not detect counterparts at optical, X-ray, or gamma-ray wavelengths.

The eclipsing redback system PSR J2055+1545 is a short-period binary and shows a large amount of orbital variability. We attempt to detect high-energy emission from the source, but observe no X-ray emission above a luminosity of roughly $4 \times 10^{30}$ erg s$^{-1}$. J2055 is also closely colocated with a Fermi source, but we are unable to detect gamma-ray pulsations with our timing solutions. Longer-term timing solutions may help to constrain better gamma-ray detections, and deeper X-ray observations may reveal emission from the system, as in other redbacks.

The new discoveries in this paper include the first two binaries found by AO327 with orbital periods shorter than 1 day. The short transit times of our search observations suggest that AO327 would be especially productive in discovering highly accelerated binary systems (i.e., with short orbital periods); however, this does not appear to be the case. The remaining 40\% of unsearched AO327 data may contain more short-period binaries.

The 16 recycled pulsars found in the AO327 data thus far represent a nearby, low-luminosity portion of the pulsar population due to the low frequency of AO327 observations. Low-frequency surveys such as these provide important complements to high-frequency surveys, which can detect more distant pulsars (e.g., \citealt{pkr+19}) and targeted radio searches of Fermi sources, which can find more luminous sources.

Despite Arecibo's decommissioning in December 2020, new findings continue to come from this survey due to the remaining unprocessed search data, as well as continued follow-up of AO327 sources with other telescopes. In the 60\% of the survey scans which we have processed, we have discovered 96 pulsars including 10 MSPs. Given this detection rate, we can expect to discover roughly 64 pulsars including 6 MSPs in the remaining 40\% of the observations to process.

\section*{Acknowledgments}

We thank the anonymous referee for comments that improved the manuscript. E.F.L., T.E.E.O., and M.A.M. are supported by NSF award AST-2009425. J.S.D. is supported by NSF award AST-2009335. We thank Jayanta Roy, Shyam Sunder, Bhaswati Bhattacharyya, and Mayuresh Surnis for their help in coordinating the GMRT observations and analyzing the data. We would also like to thank Harsha Blumer for assistance with the X-ray analysis, and Samuel Swihart for providing the information and data used to create Figure~\ref{fig:OPCM}.

The Arecibo Observatory was a facility of the National Science Foundation operated under cooperative agreement by the University of Central Florida and in alliance with Universidad Ana G. Mendez, and Yang Enterprises, Inc. We thank the staff of the GMRT who have made these observations possible. The GMRT is run by the National Centre for Radio Astrophysics of the Tata Institute of Fundamental Research, India. 

This research made extensive use of the NASA Astrophysics Data System\footnote{\url{https://ui.adsabs.harvard.edu/}}. This research has made use of the ``Aladin Sky Atlas" \citep{2000A&AS..143...33B} developed at CDS, Strasbourg Observatory, France. This work has made use of data from the European Space Agency (ESA) mission Gaia (\url{https://www. cosmos.esa.int/gaia}), processed by the Gaia Data Processing and Analysis Consortium (DPAC; \url{https://www.cosmos.esa. int/web/gaia/dpac/consortium}). Funding for the DPAC has been provided by national institutions, in particular the institutions participating in the Gaia Multilateral Agreement. The Gaia data used in this work have been provided through IRSA \citep{IRSA_GAIA} and ESA \citep{ESA_GAIA}. The Pan-STARRS data used in this work have been provided by MAST \citep{MAST}.

\vspace{5mm}
\facilities{Arecibo (PUPPI), GMRT, Swift, Fermi}
\software{
\texttt{TEMPO} (\url{https://tempo.sourceforge.net/}), 
\texttt{TEMPO2} \citep[\url{https://www.atnf.csiro.au/research/pulsar/tempo2/};][]{hem+06, tempo2}, 
\texttt{PRESTO} \citep[\url{https://www.cv.nrao.edu/~sransom/presto/};][]{smr01}, 
\texttt{PSRCHIVE} \citep[\url{https://psrchive.sourceforge.net/};][]{hsm+04, psrchive},
\texttt{DRACULA} \citep[\url{https://github.com/pfreire163/Dracula};][]{DRACULA},
\texttt{spider\_twister} \citep[\url{https://github.com/alex88ridolfi/SPIDER\_TWISTER};][]{spidertwister}
}

\bibliography{biblio}{}
\bibliographystyle{aasjournal}

\end{document}